\documentclass[twocolumn,amsmath,amssymb,superscriptaddress]{revtex4} 
\usepackage{epsfig,amsmath}

\newcommand{\beq}{\begin{equation}}
\newcommand{\eeq}{\end{equation}}
\newcommand{\beqarr}{\begin{eqnarray}}
\newcommand{\eeqarr}{\end{eqnarray}}

\begin{document}

\title{Synchronization Transition of Identical Phase Oscillators in a Directed Small-World Network}

\author{Ralf T\"onjes}
\affiliation{Ochadai Academic Production, Ochanomizu University, Tokyo 112-8610, Japan}
\author{Naoki Masuda}
\affiliation{Graduate School of Information Science and Technology, The University of Tokyo, Tokyo 113-8656, Japan}
\affiliation{PRESTO, Japan Science and Technology Agency, Kawaguchi, Saitama 332-0012, Japan}
\author{Hiroshi Kori}
\affiliation{Ochadai Academic Production, Ochanomizu University, Tokyo 112-8610, Japan}
\affiliation{PRESTO, Japan Science and Technology Agency, Kawaguchi, Saitama 332-0012, Japan}

\begin{abstract}
\noindent We numerically study a directed small-world network consisting of attractively coupled, identical phase oscillators.  While complete synchronization is always stable, it is not always reachable from random initial conditions. Depending on the shortcut density and on the asymmetry of the phase coupling function, there exists a regime of persistent chaotic dynamics. By increasing the density of shortcuts or decreasing the asymmetry of the phase coupling function, we observe a discontinuous transition in the ability of the system to synchronize.  Using a control technique, we identify the bifurcation scenario of the order parameter. We also discuss the relation between dynamics and topology and remark on the similarity of the synchronization transition to directed percolation.

\end{abstract}
\maketitle
{\bf
The adjustment of phase and frequency in large systems of oscillatory units can lead to global coherent oscillations, i.e. synchronization. On the other hand, noise and heterogeneity in the system can weaken synchronization, or even destroy it. Synchronization in the nervous system can facilitate the transfer of information or cause epileptic seizures. Multistability and hysteresis of normal and pathological collective behavior is observed. When all oscillators are identical and the coupling tends to decrease phase differences a state of complete synchronization is asymptotically stable. But even in random networks with uncorrelated and homogeneously distributed node degrees this absorbing state may not be reached or disappear when it is perturbed locally. Here we perform a detailed numerical analysis of the transition between different states of synchronization in a directed small-world network of phase oscillators. By varying the mean in-degree of the network or the nonlinearity of the phase coupling function at zero phase difference, we find discontinuous and continuous transitions with mean-field critical behavior. 
}

\section*{I. Introduction}
\noindent Synchronization in spatially extended systems and complex
networks is an important mechanism to create global spatio-temporal
correlations from local interaction rules
\cite{OsChanKu07,PiRoKurths01,Winfree80,Kuramoto84}. Its applications range from information
transfer \cite{DiAre08}, self-organized optimization of work flow or
traffic flow \cite{LaKoPeHe06} to the realization of strong coherent
oscillations in arrays of Josephson junctions or coupled fiber lasers
\cite{SiFaWie93}. The transition to partial synchronization, i.e., the
emergence of a nonzero mean-field, in systems of nonidentical
oscillators is well known.  It has been studied analytically in the
original texts by Kuramoto \cite{Kuramoto75,Kuramoto84} and in a more
general way in recent papers \cite{RespOtt05,OttAnton08,PikRos08}.
Recently it has been shown that degree heterogeneity in scale-free random
networks of identical oscillators can also lead to desynchronization or
partially synchronized states even though complete synchronization is
asymptotically stable \cite{Ermentrout08}.
\\ \\
Here we study the transitions from incoherence to partial
synchronization and to complete synchronization in a sparse, directed
small-world network of identical phase oscillators. Since the
formulation of the model \cite{WattsStrogatz98} small-world networks
have been studied as a medium for dynamical processes, such as the
spreading of epidemics \cite{NewmanWatts99}, the Ising model
\cite{OstilliMendes08,BarratWeigt00,DoGoMen02} and also synchronization
of nonidentical phase oscillators \cite{HongChoi02}.
Nontrivial behavior in sparse networks of coupled dynamical systems is
known to occur in boolean networks \cite{Derrida86,KlemmBorn05} and has
also been reported for leaky integrate-and-fire models with and without
delay coupling \cite{Roxin04,Timme08}. We show, for the case of
attractively coupled, identical phase oscillators, that even if the
in-degree distribution is homogeneous the system may enter a
quasi-stationary chaotic state of incoherence or partial
synchronization.
So far, desynchronization, due to inherent heterogeneity, could be
analytically described by global and local mean field approaches
\cite{Kuramoto84,RespOtt05,Ermentrout08}, which assume a sufficiently
large number of neighbors. In contrast, our model is highly
homogeneous, in terms of degree distribution and correlations, and none of the previously described mechanisms for desynchronization in networks of oscillators can explain the observed transition.
\begin{figure}[th!]
\center
\includegraphics[width=8cm]{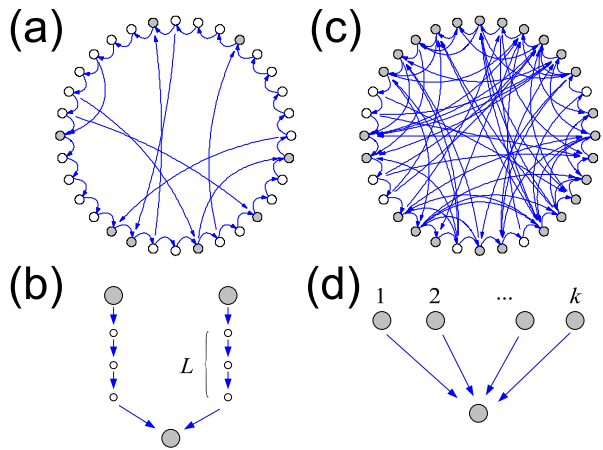}
\caption{\small The network model. Unidirectional small-world networks with $N=32$ nodes at (a), (b) low shortcut density $\sigma=0.25$ and (c), (d) higher shortcut density $\sigma=2.0$ . Joints of the network, i.e. nodes that receive more than one input, are marked gray. (b) At low shortcut densities most joints couple indirectly to two other joints through linear chain segments of length $L\approx 1/\sigma$. (d) At high shortcut densities each joint couples to $k=\sigma+1$ neighbors which are with high probability also joints.}
\label{Fig:NetworkModel}
\end{figure}
\section*{II. The Model}
The phase dynamics of $N$ identical, weakly coupled limit-cycle oscillators may be described by the Kuramoto phase equations \cite{PiRoKurths01,Kuramoto84,Kuramoto75}. In a co-rotating frame of reference where the common oscillator frequency is zero, these are
\beq	\label{Eq:KPE01}
	{\dot\vartheta}_i = \sum_{j=1}^{N} H_{ij}~g(\vartheta_j-\vartheta_i)	,
\eeq
where $H_{ij} \ge 0$ is the coupling matrix and
$g(\vartheta_j-\vartheta_i)$ is the coupling function. The coupling
function only depends on the phase difference between the oscillators
and its shape is characteristic for given coupled limit-cycle
oscillators \cite{Kuramoto84}.  A typical coupling function, which is, for
example, realized in diffusive coupling, has the properties $g(0)=0$
and $g'(0)>0$, implying that the coupling force between two coupled oscillators is attracting and vanishes when they have identical phases.
For weakly anharmonic oscillators the coupling function is usually well
approximated by the first harmonics \mbox{(See, e.g., \cite{KissHudson05})}, i.e., 
\beq	\label{Eq:CoupFkt}
	g(\Delta\vartheta) = \sin(\Delta\vartheta-\alpha) + \sin(\alpha)	.
\eeq
The parameter $\alpha$ $\left(-\pi/2<\alpha<\pi/2\right)$ breaks
the anti-symmetry of the coupling function around zero, which can lead
to nontrivial effects  \cite{Kuramoto84,Ermentrout08,BlaToe05,KuraBot02,AbraStrog04}.  
We can restrict ourselves to nonnegative values of $\alpha$
($0\le\alpha<\pi/2$) since a transformation $\alpha\to
-\alpha$ and $\vartheta_i\to -\vartheta_i$ for all $i$ leaves the
Eqs.~(\ref{Eq:KPE01}) and (\ref{Eq:CoupFkt}) invariant.
\\ \\
Under very general conditions,
i.e. nonnegative coupling strengths $H_{ij}\ge 0$ and a nondegenerate zero
eigenvalue of the coupling Laplacian, complete synchronization with
$\vartheta_i=\vartheta_j$ for all oscillators $i$ and $j$
is an asymptotically stable solution of
Eq.~(\ref{Eq:KPE01}) \cite{BlaToe09}. Strong connectedness of the network
is a sufficient condition for this. Failure of a system of identical
phase oscillators to synchronize can not be deduced from a local stability analysis of the completely synchronized state. It is the result of a
chaotic phase dynamics that riddles the phase space so that complete synchronization cannot be reached.
\\ \\
As a coupling network, we employ a well
established modification \cite{NewmanWatts99,Roxin04,BaraPecora02} of
the original Watts Strogatz model \cite{WattsStrogatz98}.
Starting with
a ring of $N$ unidirectionally coupled oscillators
we add $N_\textnormal{sc}$
unidirectional shortcuts with random origin $i$ and destination $j$ (See
Fig.~\ref{Fig:NetworkModel}). We refer to nodes which receive more than one input as the joints of the network.
In addition to the system parameter $\alpha$,
which characterizes the phase coupling function, the shortcut density
$\sigma = N_\textnormal{sc}/N$ characterizes the topology of the
coupling network.
\\ \\
In the present paper, except for Fig.~\ref{Fig:sigma_alpha_scans}b,
we assume uniform and normalized input strength, i.e.,
\beq \label{Eq:InputNorm}
	\sum_{j=1}^{N}H_{ij} = 1
\eeq
for each oscillator and $H_{ij}=1/k_i^{\textnormal{in}}$ if oscillator $i$ with in-degree $k_i^{\textnormal{in}}$ couples to $j$ or $H_{ij}=0$ otherwise.
The normalization would have a great impact on the dynamics if the
coupling network was very heterogeneous. In the nonnormalized case and when the phases are uniformly distributed the phase velocity of each oscillator is biased proportional to its in-degree and $\sin\alpha$ \cite{Ermentrout08}. Oscillators with larger in-degree tend to move faster so that degree heterogeneity indirectly leads to a heterogeneity in frequencies. In contrast, when the coupling strength is normalized to unity the bias to the phase velocity is equal to $\sin\alpha$ for all oscillators (See Fig.~\ref{Fig:sigma_alpha_scans}c).
However, the directed small-world network model that we use has a Poissonian degree
distribution, and is thus a homogeneous network.  In fact, we 
observe similar synchronization behavior with and without the
normalization given by Eq.~(\ref{Eq:InputNorm}) (See
Figs.~\ref{Fig:sigma_alpha_scans}a and \ref{Fig:sigma_alpha_scans}b).
Another consequence of this normalization is that the phase velocities are
bounded, i.e., $-1\le{\dot\vartheta}_i-\sin\alpha\le 1$ and an
oscillator that receives only a single input is always phase
locked to it. Phase slips do not occur in a chain of
unidirectionally coupled phase oscillators but only at the joints of a network.
\section*{III. Simulation Results}
\begin{figure}[!t] 
\center
 \includegraphics[width=4.1cm]{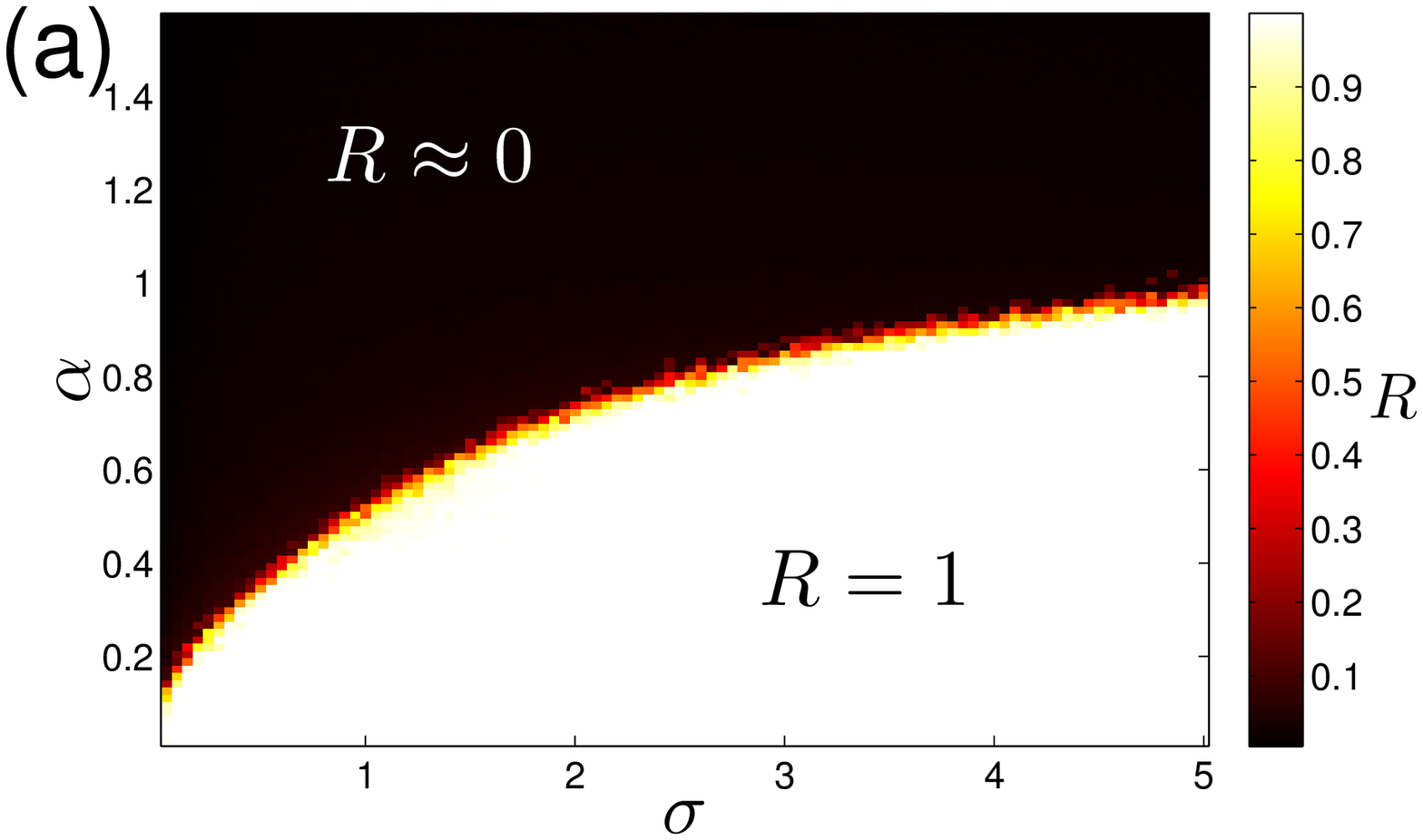}
 \includegraphics[width=4.1cm]{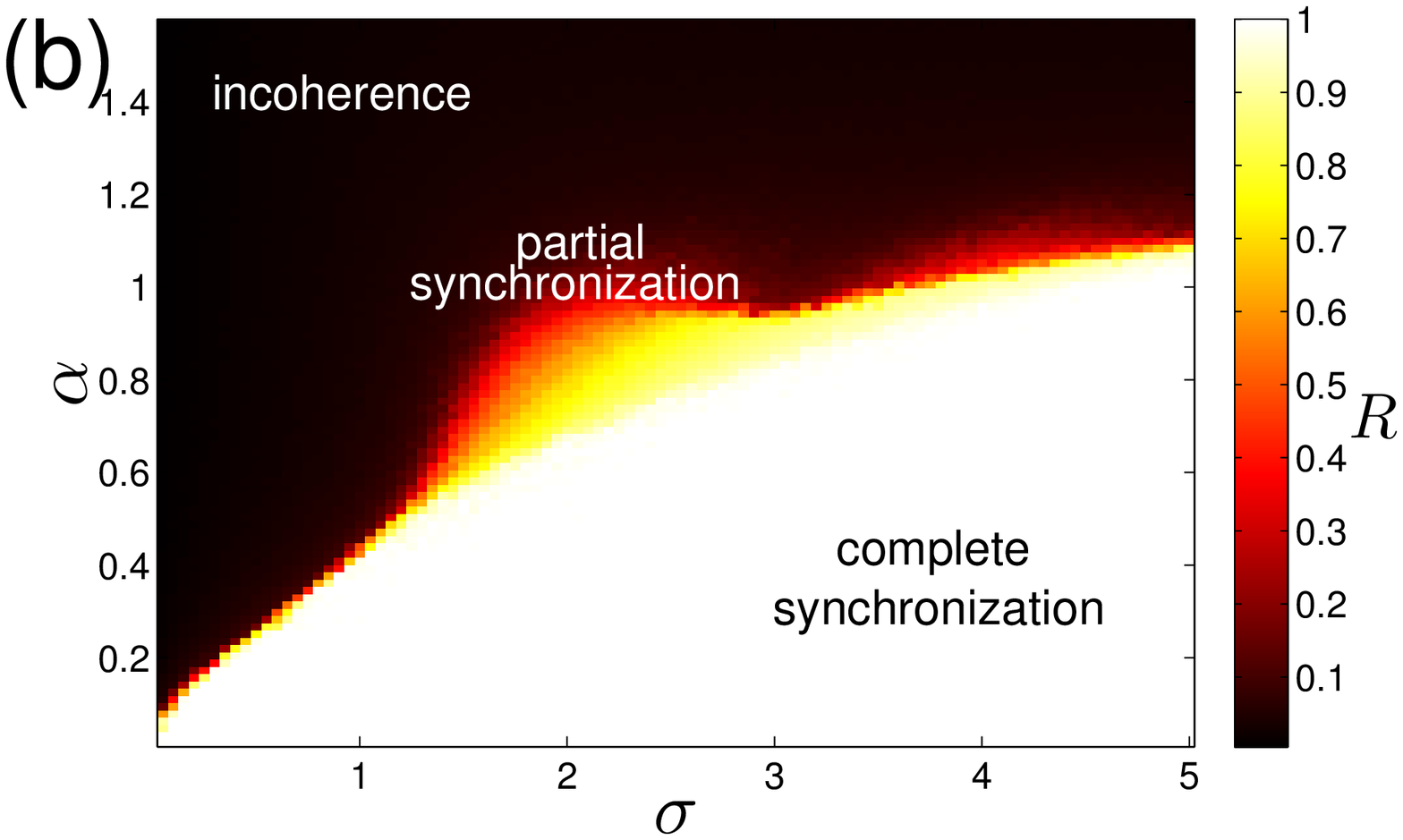}
 \includegraphics[width=4.1cm]{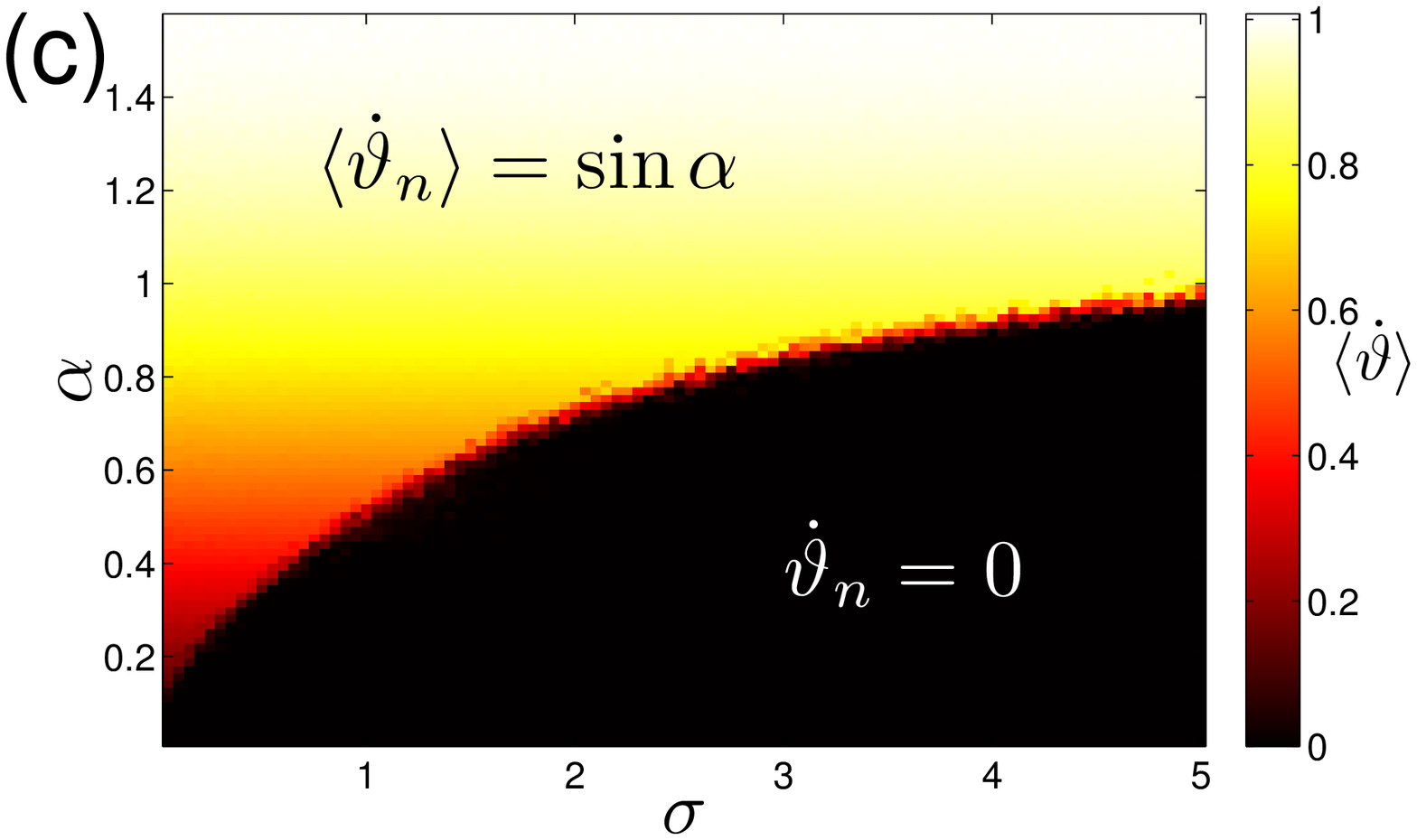}
 \includegraphics[width=4.1cm]{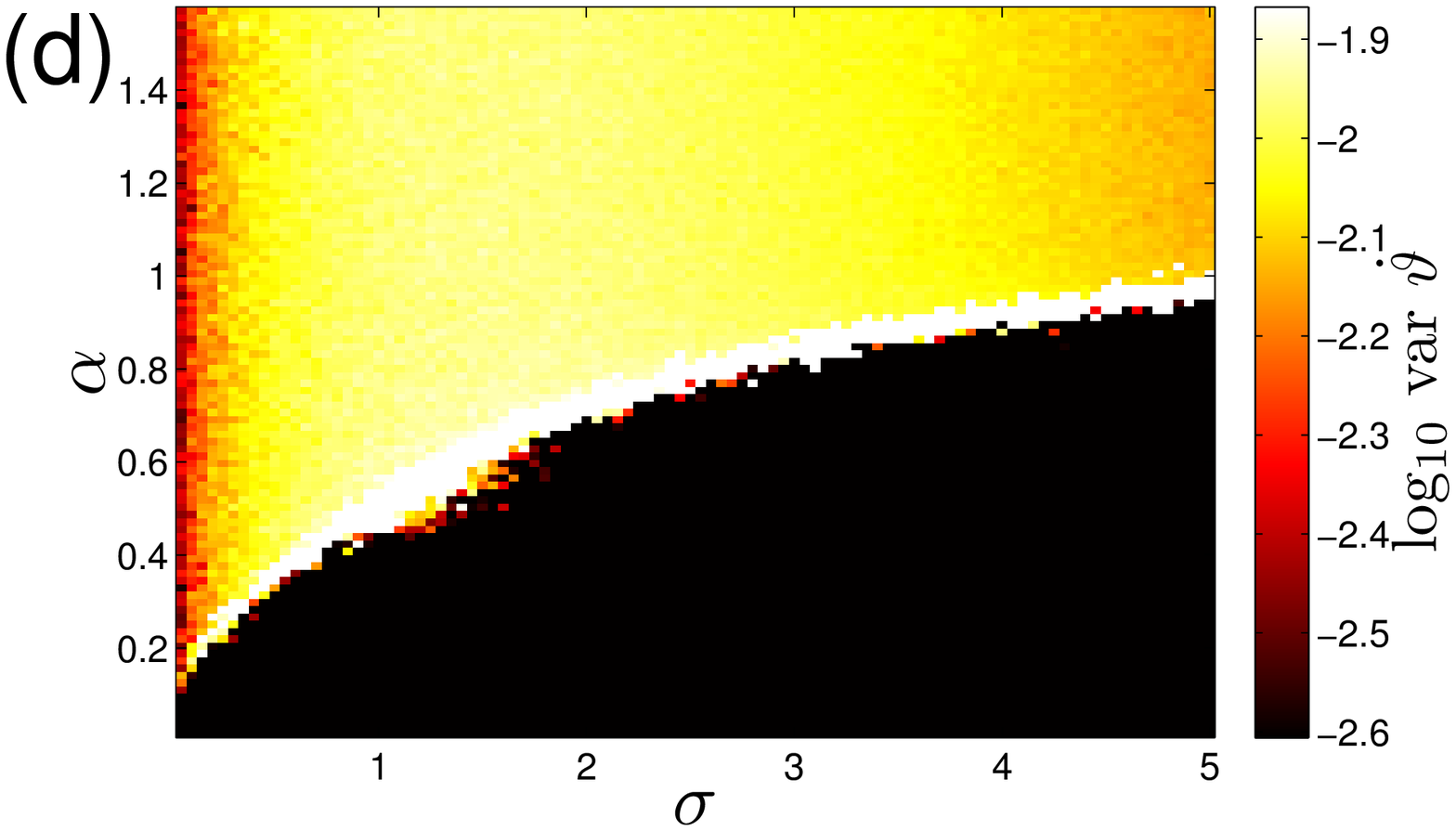}
\caption{\small Synchronization transition in the $(\sigma,\alpha)$ parameter plane. Each point corresponds to an ensemble average over 10 network realizations ($N=800$) and time average over 600 units after an initial transient of 200. Shown are (a) the mean order parameter $R$, (c) the mean oscillator frequency and (d) the variance of phase velocities in the case of normalized input strength. For comparison we also show (b) the mean order parameter for nonnormalized coupling strength for which a larger area of partial synchronization is observed at intermediate shortcut densities.}
\label{Fig:sigma_alpha_scans}
\end{figure}

Figure \ref{Fig:sigma_alpha_scans} shows the phase diagrams obtained from numerical integrations of Eq.~(\ref{Eq:KPE01}) for networks of
$N=800$ oscillators starting from uniformly random initial conditions. For each set of parameter values for $\sigma$ and
$\alpha$, we show time and ensemble averages of various quantities. Numerical
simulations of an ensemble of 10 network realizations were run for $T=800$ where an initial transient time of $200$ time units was disregarded.  In
Figs.~\ref{Fig:sigma_alpha_scans}a and \ref{Fig:sigma_alpha_scans}b, the average order parameter $R=\langle r(t) \rangle_{\rm
time, trials}$, where $r(t)=\left|\frac{1}{N}\sum_i
e^{\textnormal{i}\vartheta_i(t)}\right|$ is displayed.  Note that $r(t)=1$ for
complete synchronization and $R \sim O(1/\sqrt{N})$ for complete incoherence, i.e.,
uniform phase distribution. There exists a clear transition
from an incoherent regime $(R\approx 0)$ at small shortcut densities
$\sigma$ or larger asymmetry $\alpha$ to a coherent regime $(R=1)$ at
higher shortcut density or lower asymmetry.  
To quantify the dynamical properties of the incoherent state, we also observed the mean frequency
$\langle {\dot\vartheta}_i(t) \rangle_{i,~\rm time,~trials}$
(Fig.~\ref{Fig:sigma_alpha_scans}c) and the variance of the phase
velocities, $\langle {\rm var}\ {\dot\vartheta}_i(t) \rangle_{i~\rm time,
~trials}$ (Fig.~\ref{Fig:sigma_alpha_scans}d). 
\begin{figure}[!t] 
\center
 \includegraphics[width=4.1cm]{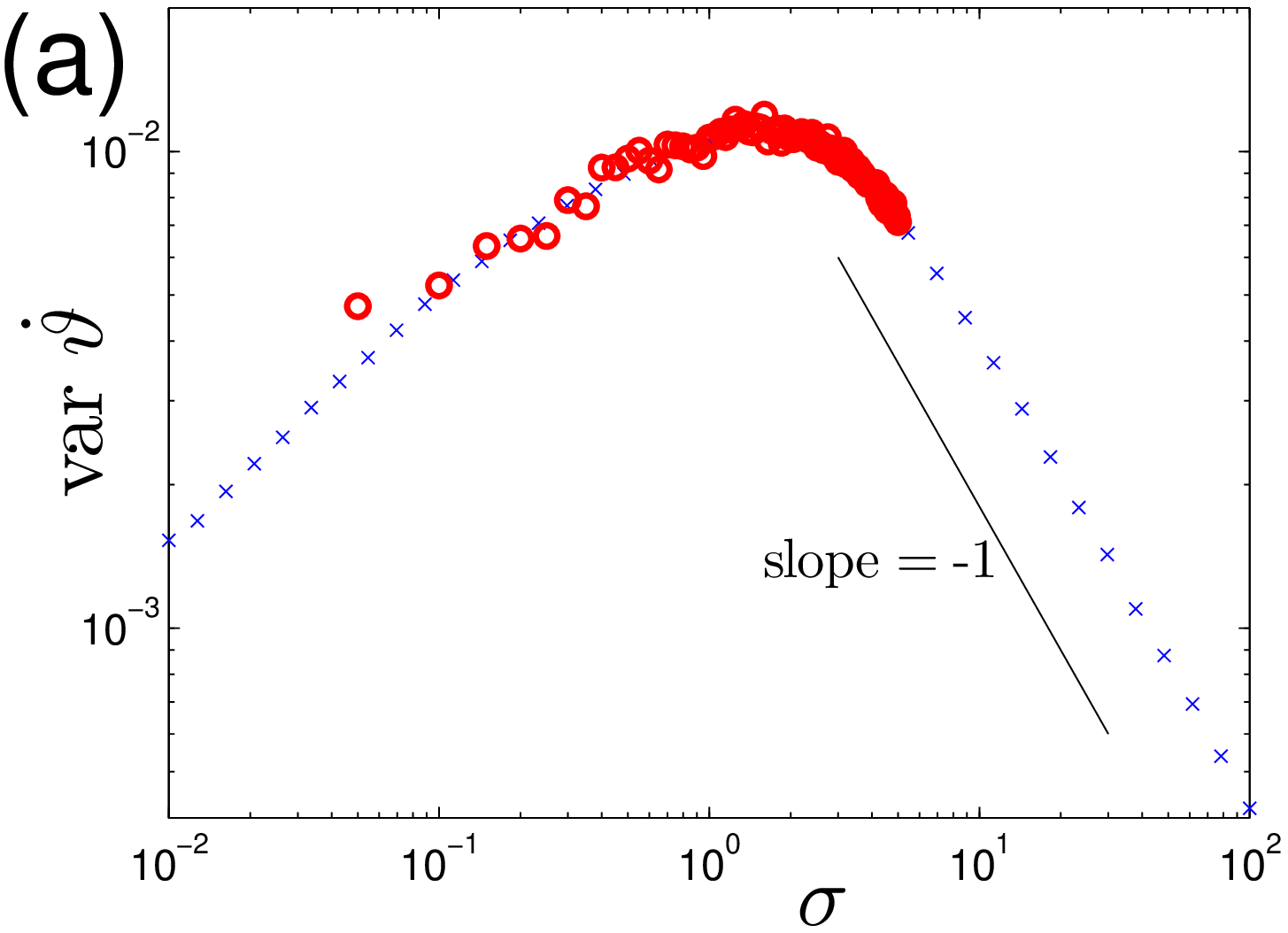}
 \includegraphics[width=4.1cm]{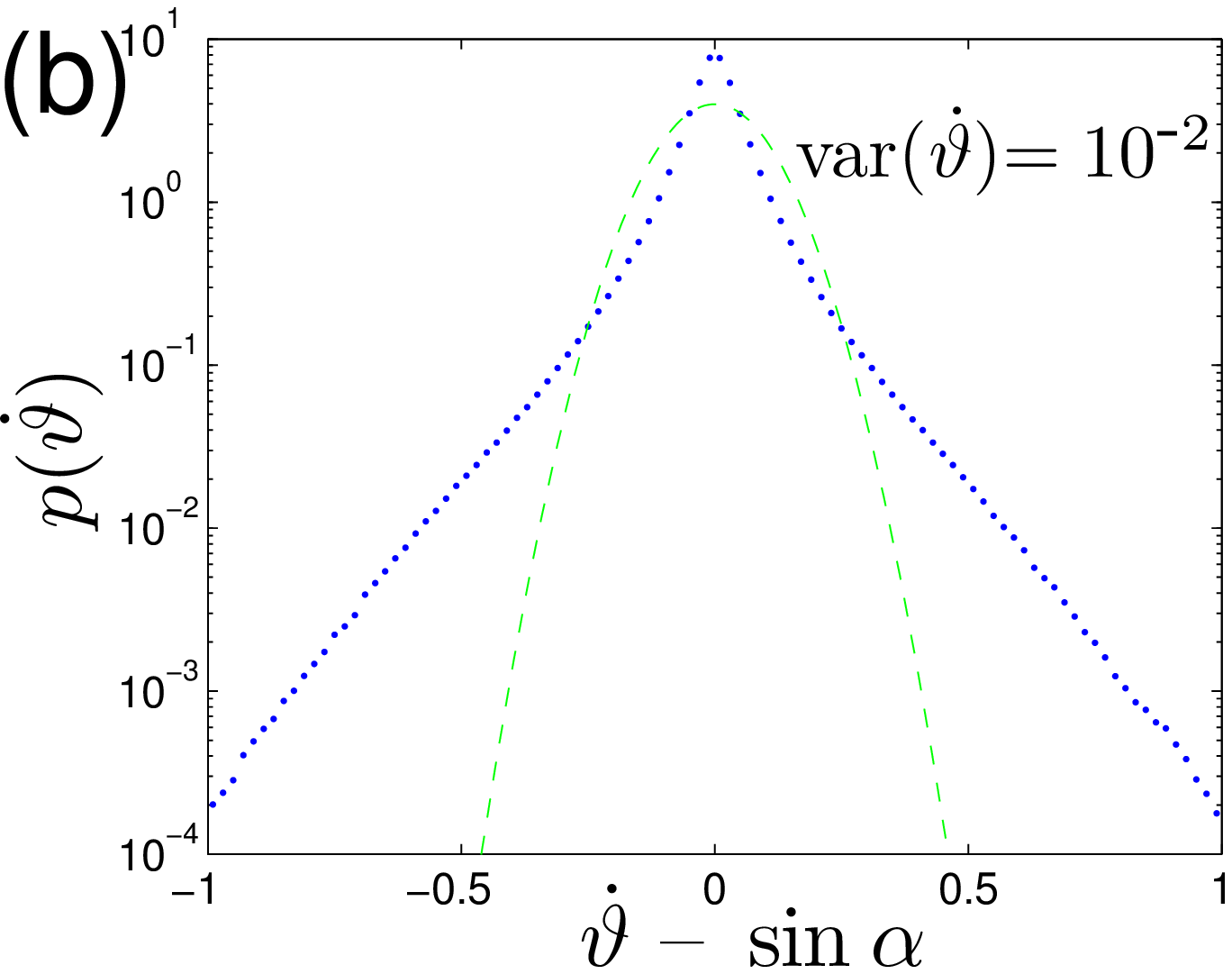}
\caption{\small Variance and distribution of phase velocities in the incoherent state. (a) Variance of phase velocities obtained from simulations with $N+N_{sc}=10^6$ and $\alpha=\pi/2$ (crosses) and from the simulations presented in Fig.~\ref{Fig:sigma_alpha_scans}d (circles) at $\alpha=1.2$. (b) The distribution of phase velocities in the incoherent state at $\sigma=1.0$ (dots) is centered around the mean of $\langle{\dot\vartheta}\rangle=\sin\alpha$. It is peaked at the center and much broader than a Gaussian distribution of the same variance (dashed line).}
\label{Fig:structural_resonance}
\end{figure}
\\ \\
\mbox{Figure \ref{Fig:structural_resonance}}b shows the distribution of phase velocities, which is characteristic for the phase diffusion process in the incoherent state.
The variance of the phase velocities is a measure for the internal noise due to chaotic phase dynamics. We observe
that the mean frequency in the incoherent state with normalized
input strength is equal to $\sin\alpha$ independently of
$\sigma$, while the variance of the phase velocities is independent of
$\alpha$.
\begin{figure}[!t] 
\center
 \includegraphics[width=4.1cm]{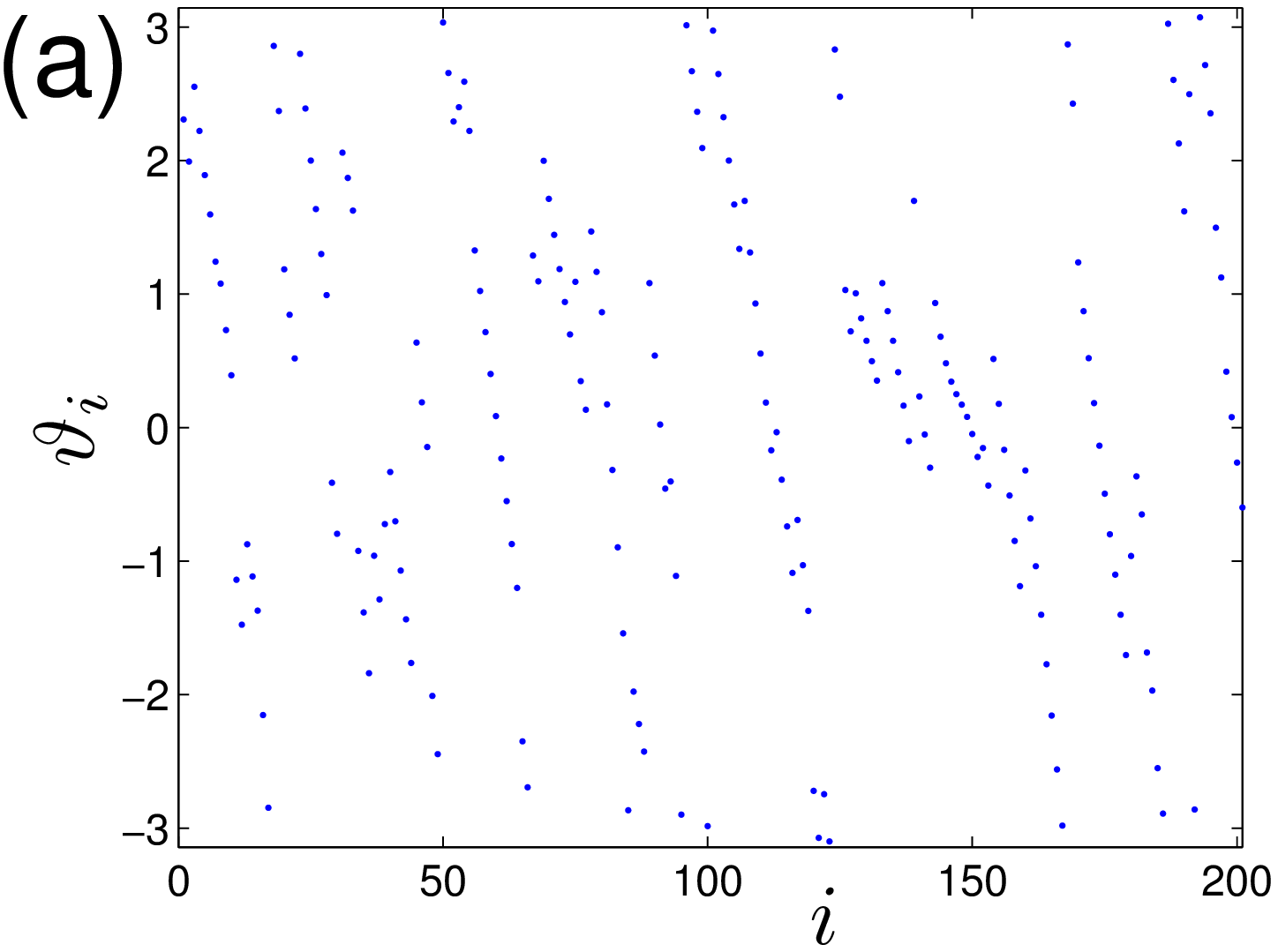}
 \includegraphics[width=4.1cm]{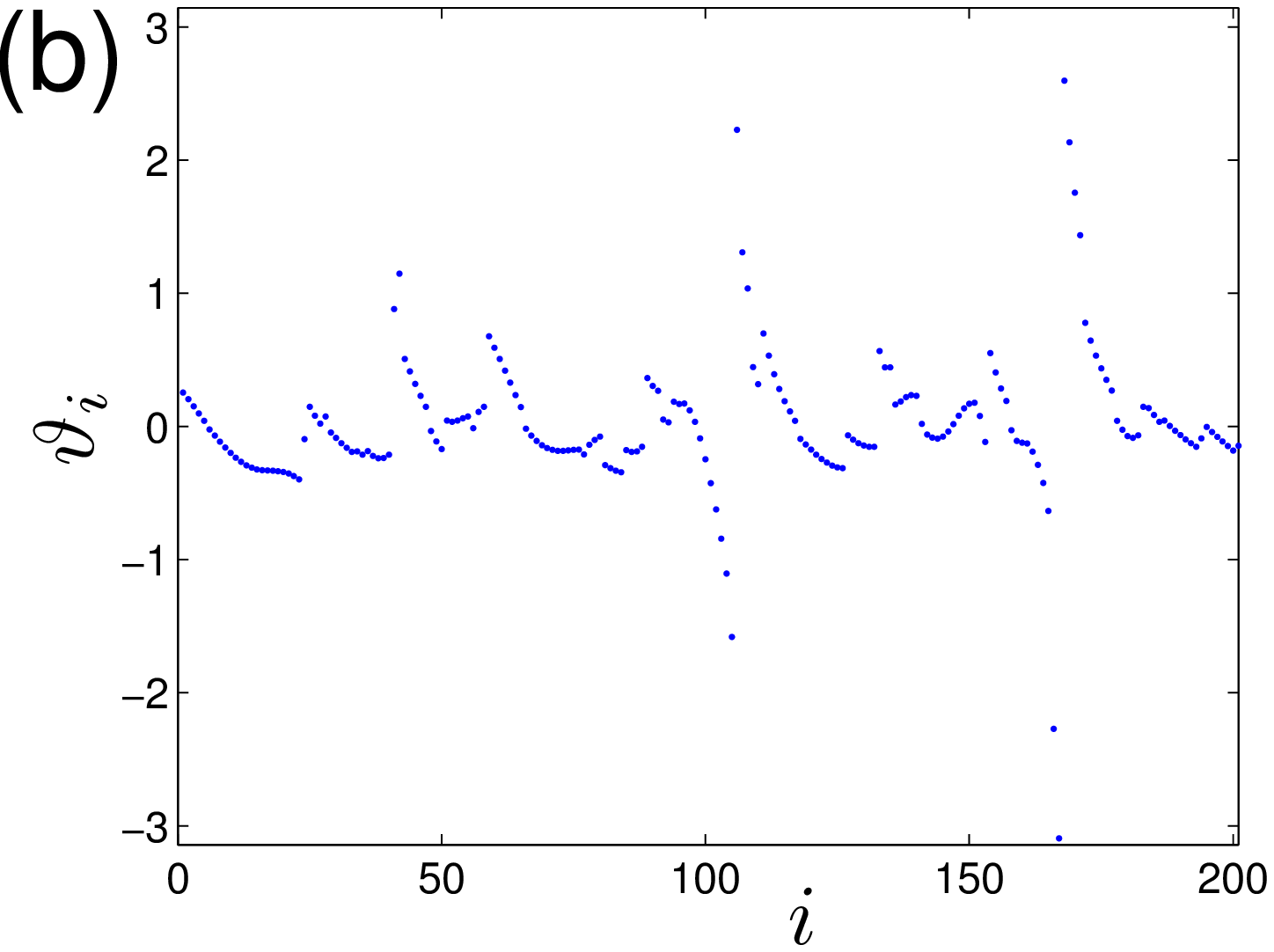}
 \includegraphics[width=4.1cm]{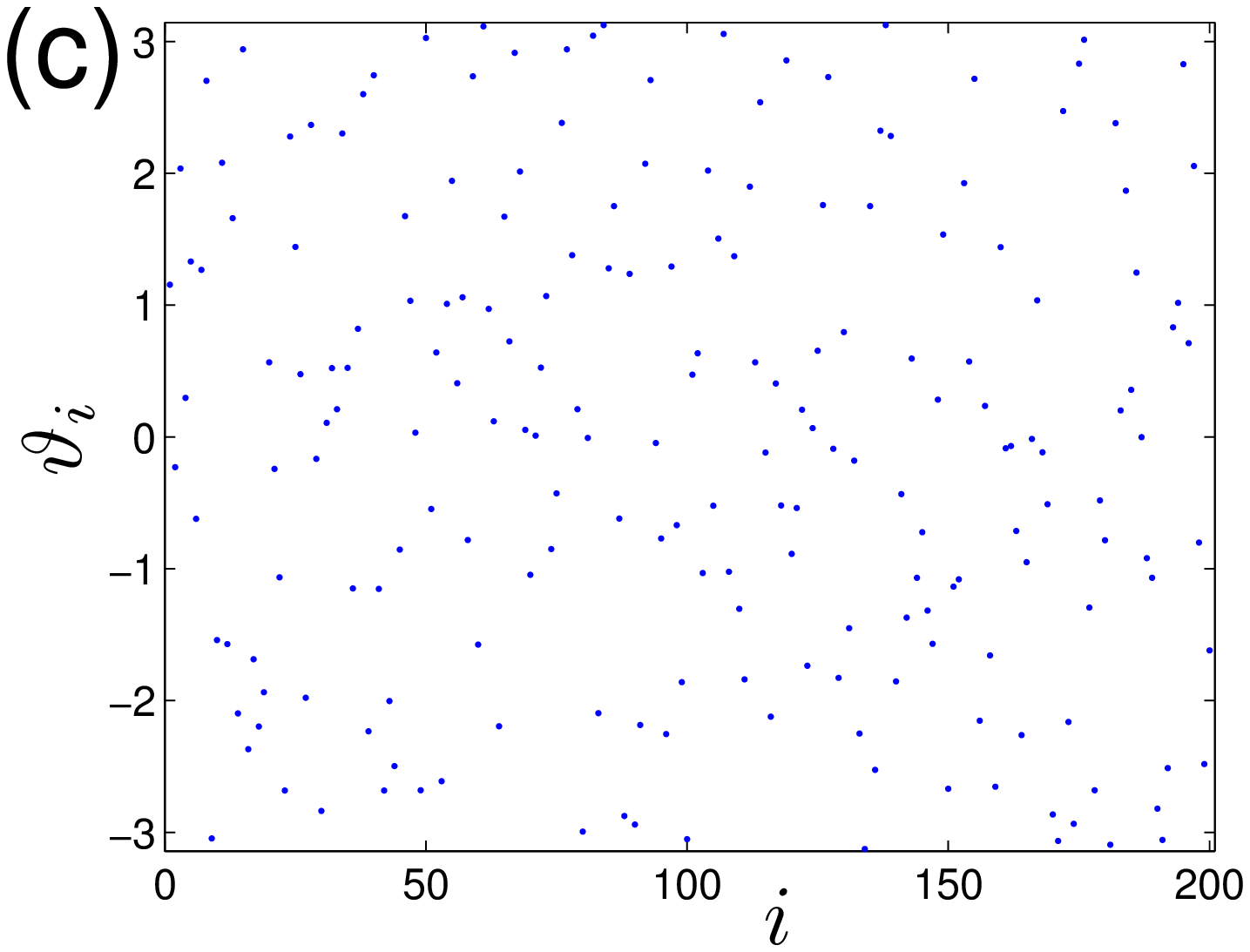}
 \includegraphics[width=4.1cm]{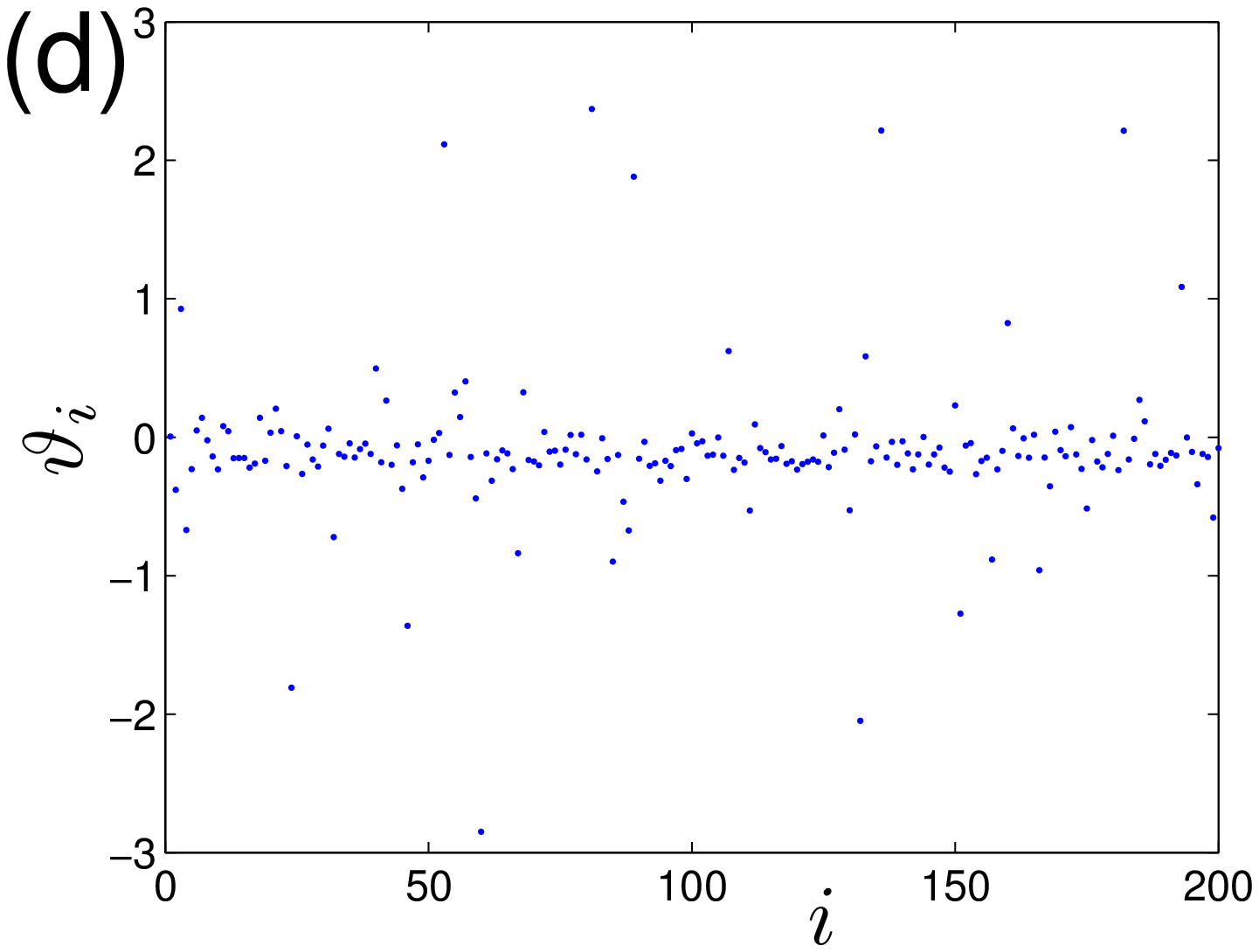}
\caption{\small Snapshots of the first $200$ phases in a system of $N=1000$ oscillators in a dynamical (quasi) equilibrium state for low shortcut density $\sigma=0.35$ ($\alpha=0.36$) in (a) and (b) and high shortcut density $\sigma=128$ ($\alpha=1.5$) in (c) and (d). Subfigures (a) and (c) show the stable incoherent state in a parameter region of bistability with the partially synchronized states shown in (b) and (d). Only at low shortcut densities the phases have a spatio-temporal structure at the length scale $1/\sigma$ of the chain segments.}
\label{Fig:phase_snapshot}
\end{figure}
\\ \\
Figure \ref{Fig:phase_snapshot} shows snapshots of the
first $200$ phases in a network of $N=1000$ oscillators.  In
the incoherent state for low $\sigma$ (Fig.~\ref{Fig:phase_snapshot}a),
the one dimensional chain segments sustain traveling waves with an
average phase difference of $\alpha$ between neighboring oscillators. 
Forward and backward phase slips occur occasionally in the joints of the network when the phase of the local mean field changes rapidly as it passes the vicinity of zero.
In the partially synchronized state for low $\sigma$
(Fig.~\ref{Fig:phase_snapshot}b), the phases of the chain segments
evolve around a global mean field. When a forward slip between the phase of a joint and the global mean field occurs, all oscillators in the chain segment adjacent to this joint will also perform a phase slip at a delayed time.
For a high shortcut density no spatial patterns exist (Fig.~\ref{Fig:phase_snapshot}c, d).
\subsection*{Control scheme and bifurcation scenario}
A more detailed examination using a linear control scheme reveals the
bifurcation scenario for the order parameter.  Assuming that the order
parameter $r(t)$ evolves according to some unknown mean field dynamics,
we make the ansatz
\beq
	\dot r = f(r,\alpha,\sigma,\eta(t))	,
\eeq
where $\eta(t)$ represents intrinsic or external noise. In the absence of
noise the dynamics under the general linear control scheme
\beq
	\dot\alpha =  c_0 (r-r_0) + c_1\dot r 
\eeq
\begin{figure}[!t] 
\center
 \includegraphics[width=4.1cm]{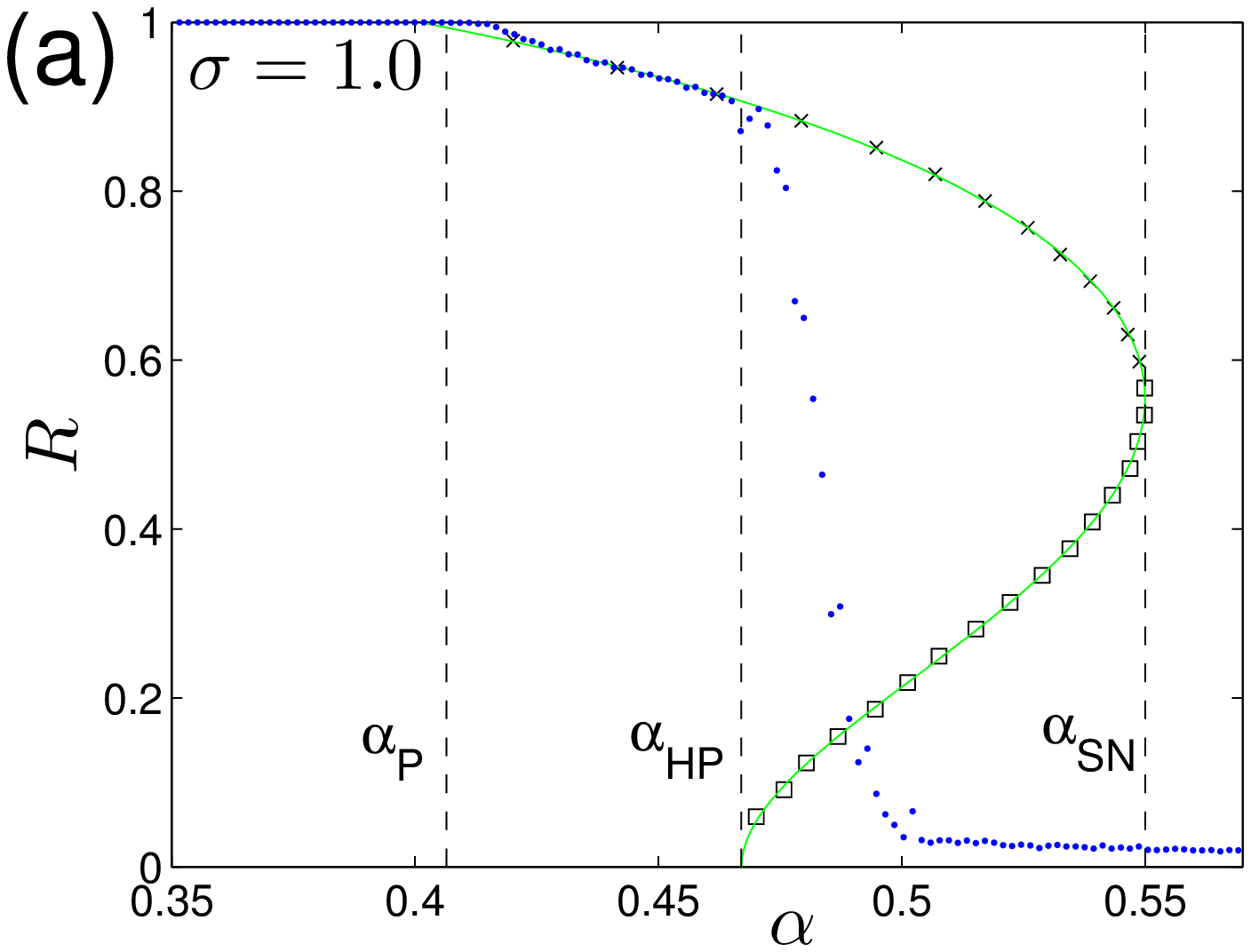}
 \includegraphics[width=4.1cm]{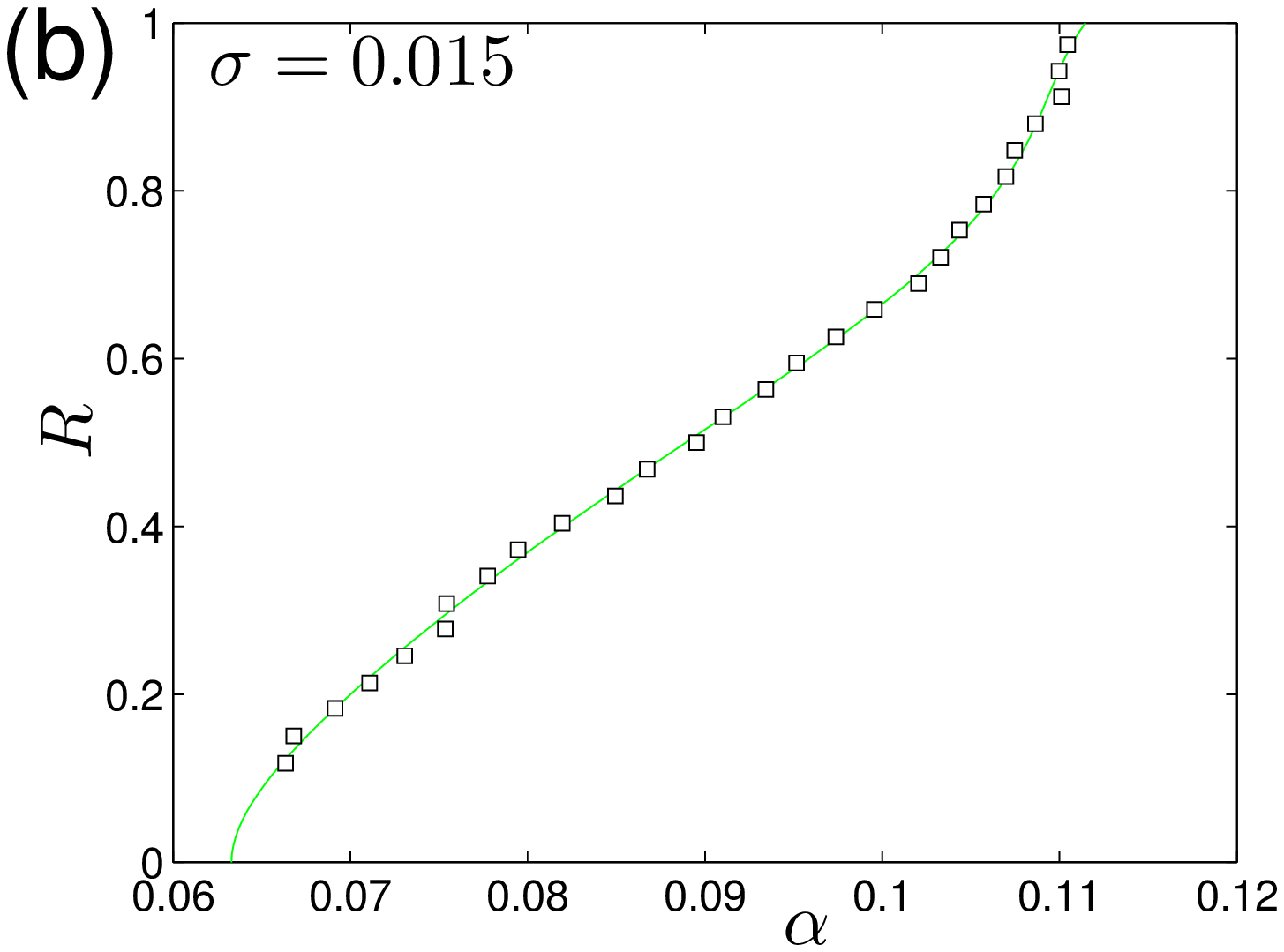}
 \includegraphics[width=4.1cm]{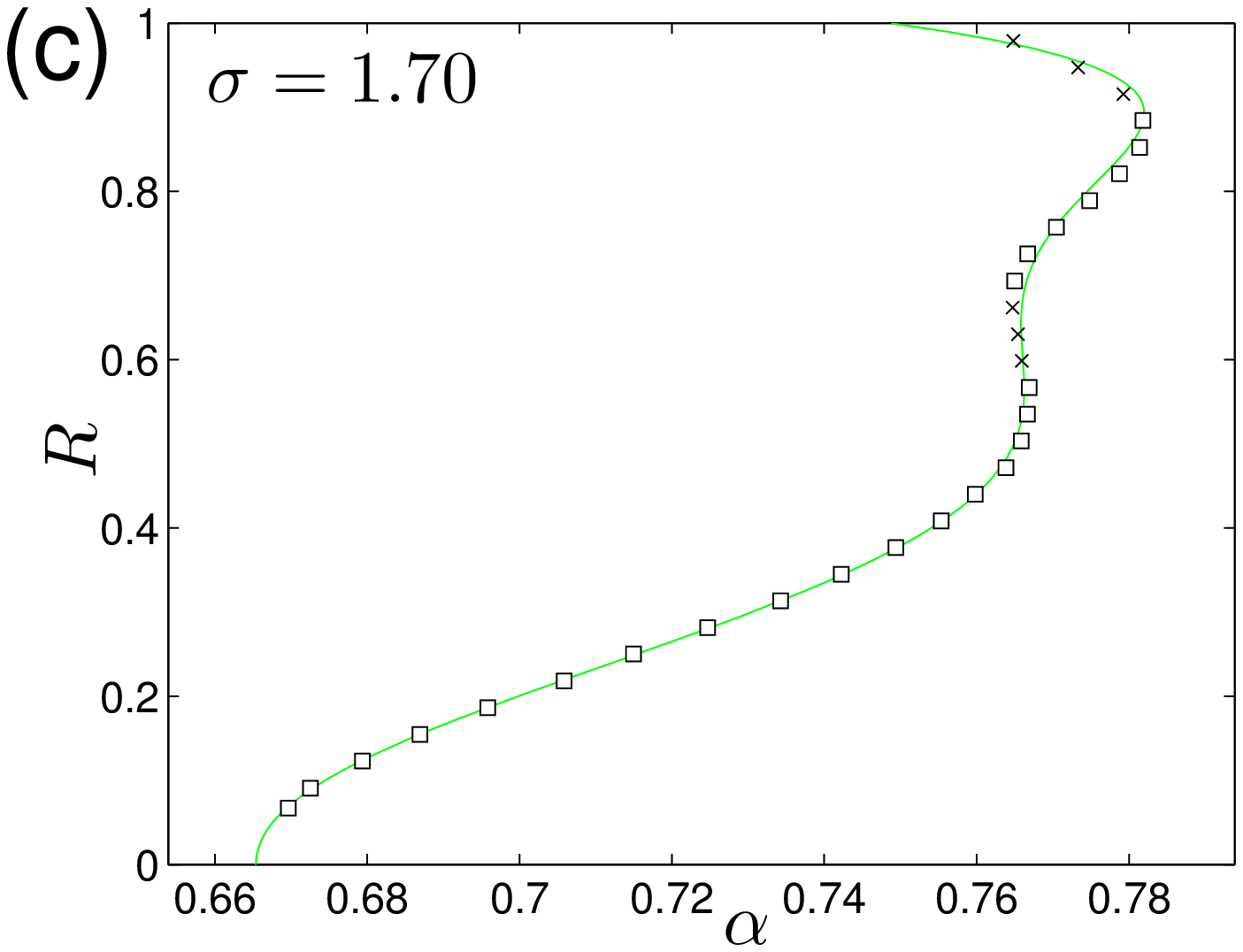}
 \includegraphics[width=4.1cm]{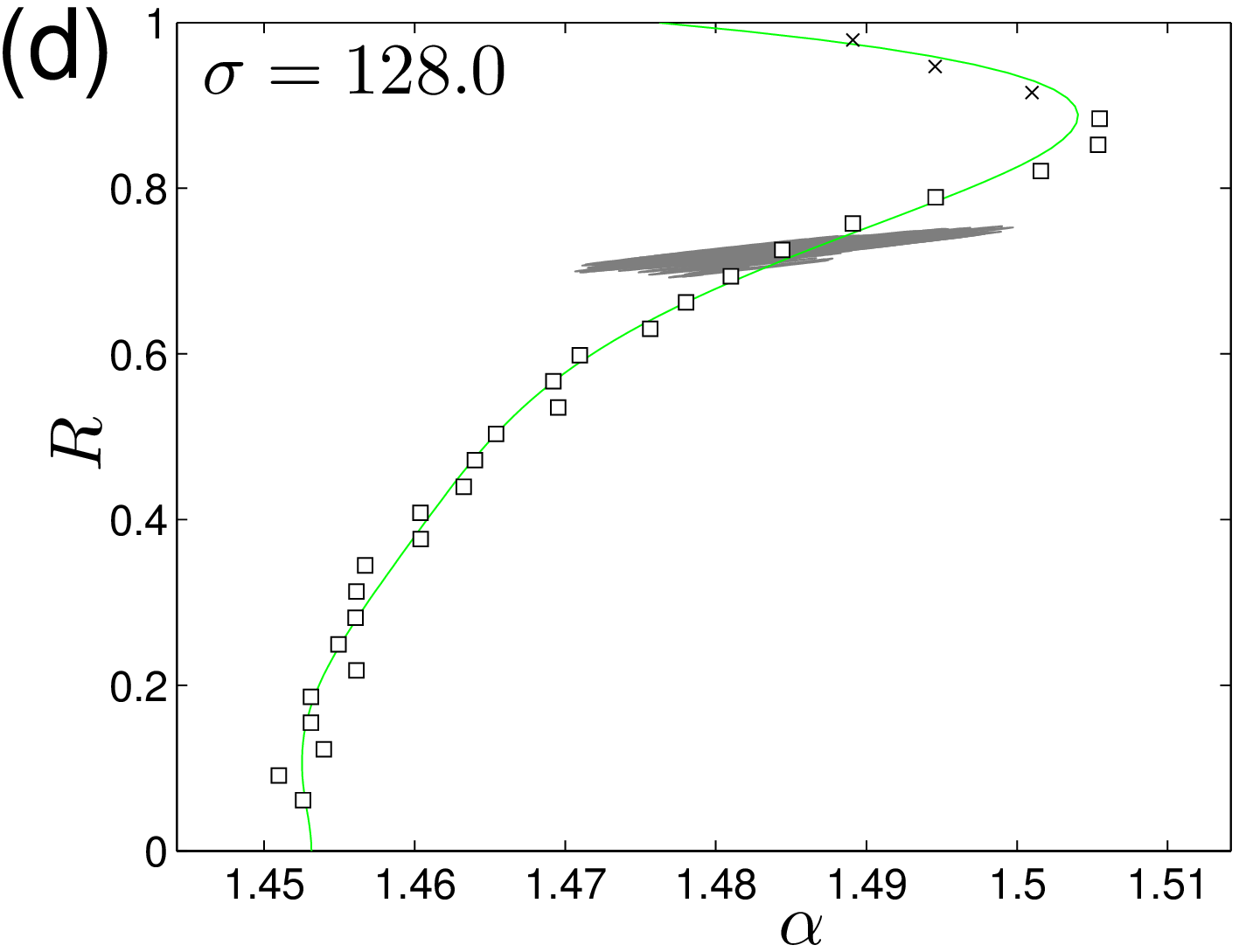}
\caption{\small Bifurcation diagram of the order parameter $R$ as a function of control parameter $\alpha$ at selected values of shortcut densities $\sigma$. Points on the branches of unstable (open squares) and stable (crosses) partially synchronized states were obtained as averages of the trajectory $(R,\alpha)$ (see light gray area in (d)) under the control scheme given by Eq.~(\ref{Eq:AvgControl}). The green lines are sixth order polynomial fits $\alpha(R)$ constrained to $\alpha'(0)=0$ because of the assumption of a Hopf bifurcation of the incoherent state. From these fits we also find the threshold $\alpha_{\textnormal{P}}=\alpha(1)$ for complete synchronization and the points of saddle node bifurcations $\alpha_{\textnormal{SN}}$ of stable and unstable partially synchronized states. The dots in (a) are the average order parameter in simulation with networks of $N=8000$ oscillators. Each point is an ensemble average of the order parameter over 50 realizations after 1000 units of time.}
 \label{Fig:rcontrol}
\end{figure}
%
\noindent
has the fixed point $r=r_0$ and $\alpha=\alpha_0(r_0,\sigma)$ with $\dot
r = f(r_0,\alpha_0,\sigma,0) = 0$. 
In Appendix B we show that, in the absence of noise, sufficiently large positive values of $c_0$ and $c_1$ can stabilize any fixed point $r_0$.
In the presence of noise, the time derivative of $r$ may not be
well defined. We thus use a short-time average of $r$ and $\dot r$ instead of their instantaneous values, i.e.,
\beq	\label{Eq:AvgControl}
	\dot\alpha =  c_0 \left(\left\langle r(t) \right\rangle_\gamma -
	r_0 \right) + c_1 \left\langle{\dot r}(t)\right\rangle_\gamma		,
\eeq
where
\beqarr	\label{Eq:LaplaceR}
	\left\langle r(t) \right\rangle_\gamma &=& \gamma
	\int\limits_{0}^{\infty} r(t-\tau) e^{-\gamma \tau} d\tau , \qquad\textnormal{and}	\nonumber  \\ \\
	\left\langle \dot r(t) \right\rangle_\gamma &=& \gamma
	\int\limits_{0}^{\infty} \dot r(t-\tau) e^{-\gamma \tau} d\tau	\nonumber
\eeqarr
\begin{figure}[!t] 
\center
 \includegraphics[width=4.1cm]{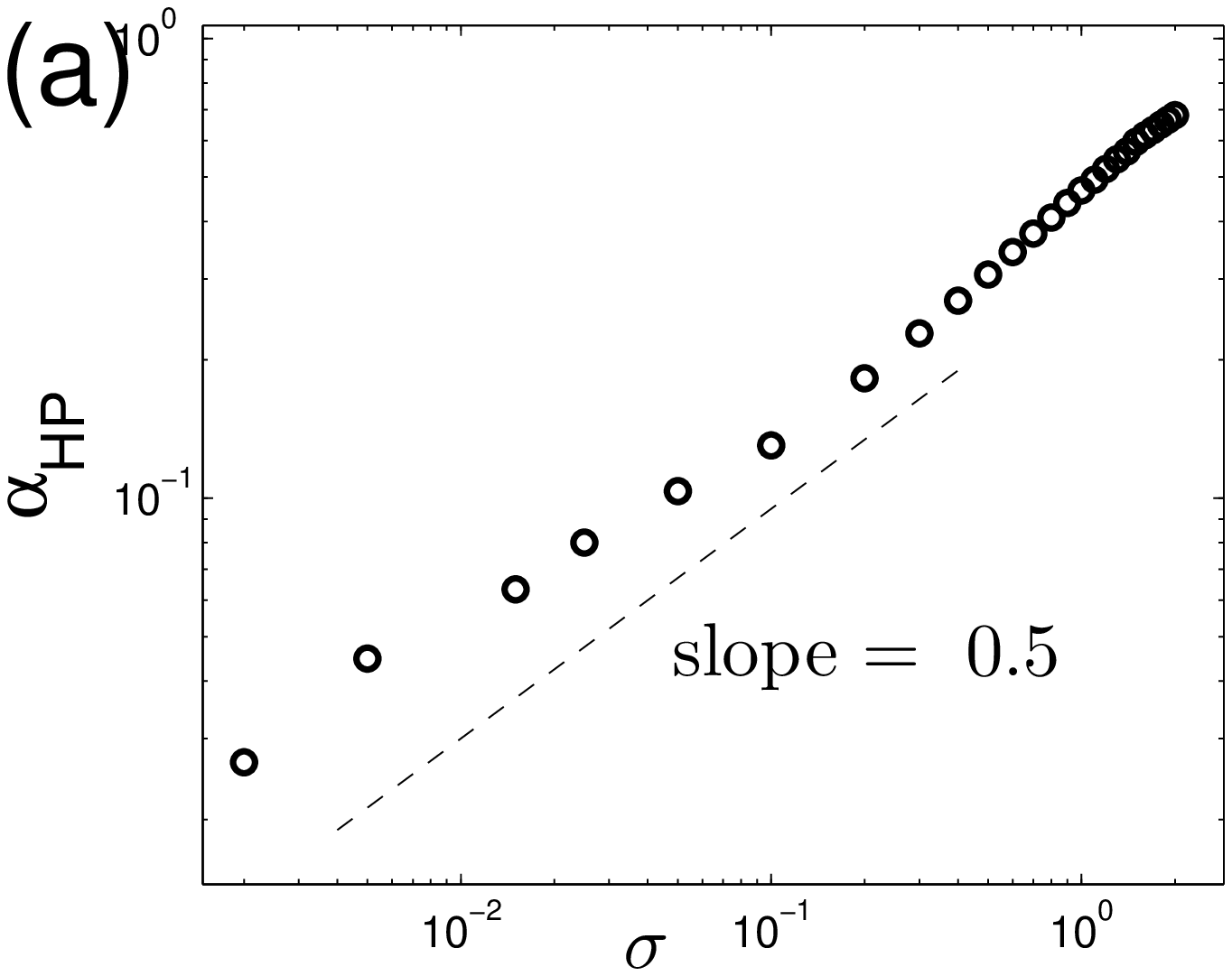}
 \includegraphics[width=4.1cm]{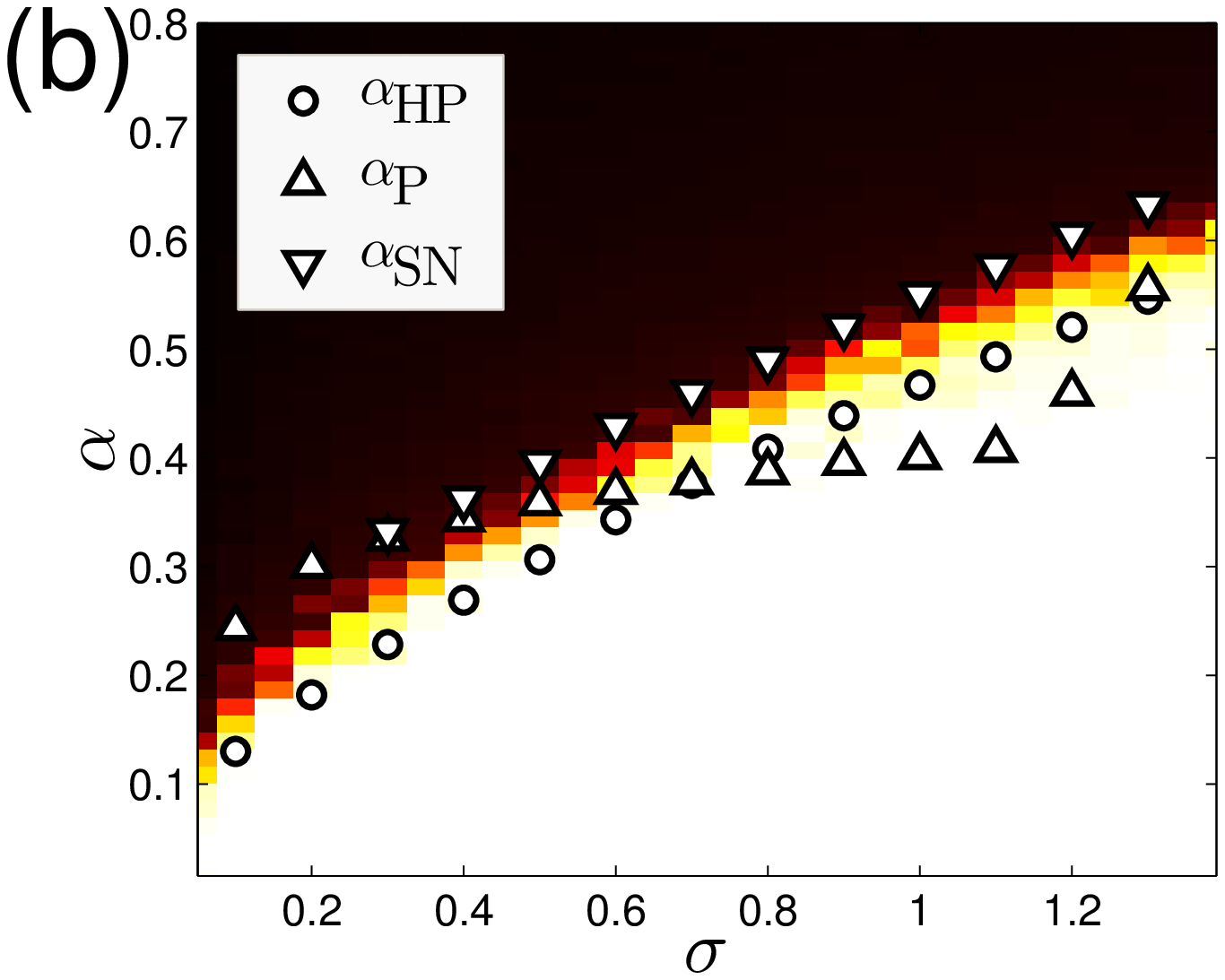}
 \includegraphics[width=4.1cm]{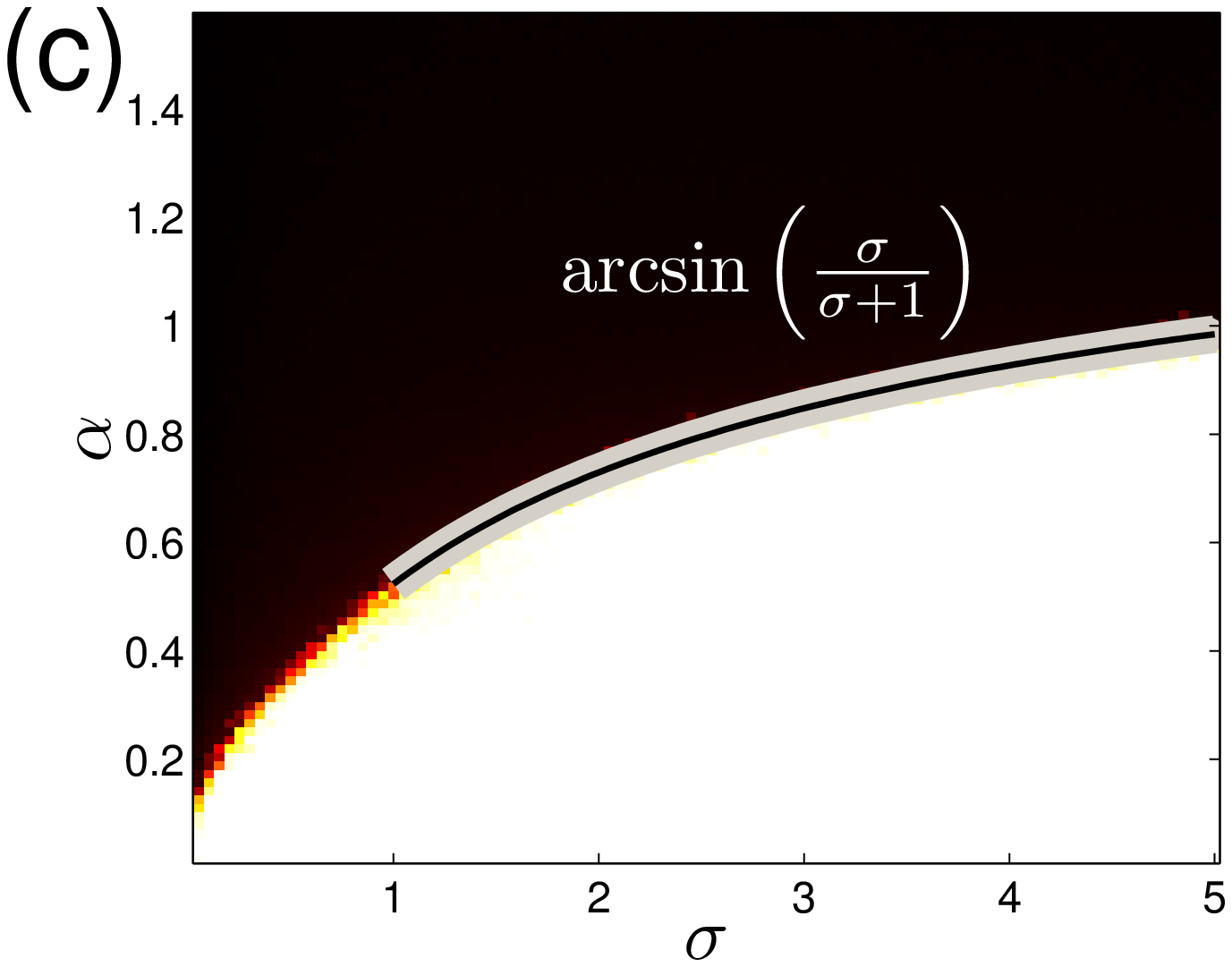}
 \includegraphics[width=4.1cm]{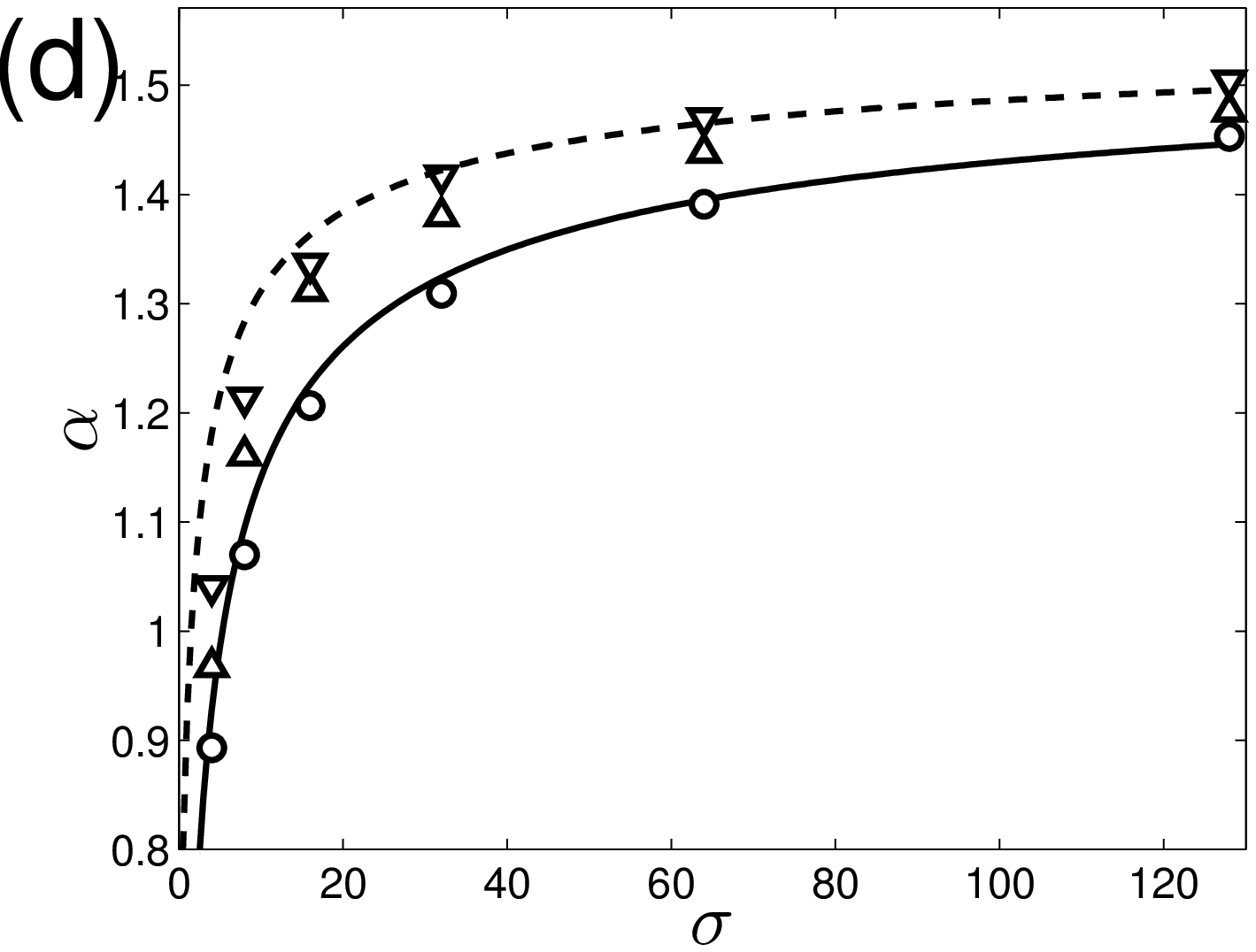}
\caption{\small Numerically determined synchronization points for (a) low shortcut densities, (b) and (c) intermediate shortcut densities and (d) large shortcut densities. The open circles, the upward and the downward triangles mark the Hopf bifurcation points $\alpha_{\textnormal{HP}}(\sigma)$ of the incoherent state, the transition points $\alpha_{\textnormal{P}}(\sigma)$ to complete synchronization and the saddle-node bifurcation points $\alpha_{\textnormal{SN}}(\sigma)$, respectively. Stable partial synchronization is found between $\alpha_{\textnormal{P}}(\sigma)$ and $\alpha_{\textnormal{SN}}(\sigma)$. At intermediate to large shortcut densities (c) and (d) the transition to synchronization is very well described by $\alpha_{\textnormal{HP}}(\sigma)=\arcsin(\sigma/(\sigma+1))$ (solid line). This is not the case for very low shortcut densities (a) where $\alpha_{\textnormal{HP}}(\sigma)$ approaches zero more slowly than linearly. The line of slope 0.5 in the double-logarithmic plot (a) is drawn for comparison. The critical line $\alpha_{\textnormal{HP}}(\sigma)=\arccos(0.85 /\sqrt{\sigma+1})$ obtained from the heuristic mean field ansatz Eq.~(\ref{Eq:MF01}) (dashed line in Subfig.~(d)) agrees qualitatively with the asymptotic approach of $\alpha_{\textnormal{HP}}$ to $\pi/2$ but is larger than the values obtained by our control scheme (open circles). The color code for the background of (b) and (c) is the same as in Fig.~\ref{Fig:sigma_alpha_scans}a and b.}
 \label{Fig:sa_bif}
\end{figure}
are interpreted as stochastic integrals. \\ \\
Using the control scheme given by Eq.~(\ref{Eq:AvgControl}), and appropriately tuned parameters $\gamma$, $c_0$ and $c_1$,
we succeeded to trace the stable and unstable branches in the
bifurcation diagram of $r$ as a function of the control parameter
$\alpha$ at fixed shortcut densities $\sigma$ (Fig.~\ref{Fig:rcontrol}). 
We find that the incoherent state always loses stability in a discontinuous transition at a critical value $\alpha_{\textnormal{HP}}(\sigma)$. We conjecture that this transition is a sub-critical Hopf-Bifurcation of the complex mean field. The branch of unstable partially synchronized states may fold back and become stable in a saddle-node bifurcation at a parameter $\alpha_{\textnormal{SN}}(\sigma)$. This hysteresis behavior is most pronounced around $\sigma\approx 1$ with possibly small parameter regions in which multiple partially synchronized states are stable (Fig.~\ref{Fig:rcontrol}c). At $\alpha_{\textnormal{P}}$ the branches of partial synchronization connect to the absorbing state of complete synchronization with $r=1$. In the next section we perform a finite size scaling analysis at this transition point and conclude that it is of mean field directed percolation universality.
From these simulations, we could determine the three curves $\alpha_{\textnormal{HP}}(\sigma)$, $\alpha_{\textnormal{SN}}(\sigma)$ and $\alpha_{\textnormal{P}}(\sigma)$ for a large range of shortcut densities (Fig.~\ref{Fig:sa_bif}).  
\subsection*{Nonequilibrium Transition to Complete Synchronization}
\begin{figure}[!tbh] 
\center
 \includegraphics[width=4.1cm]{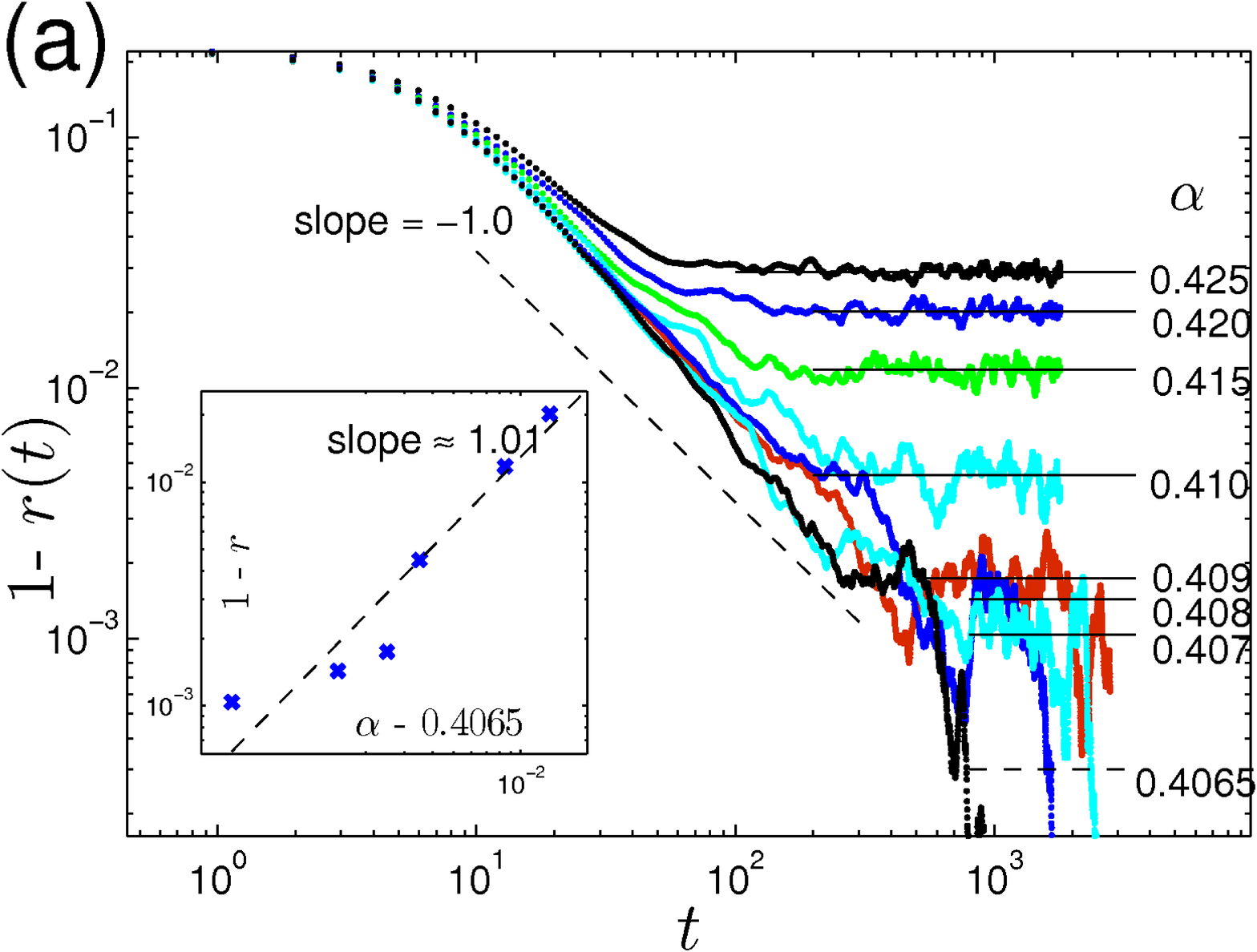}
 \includegraphics[width=4.1cm]{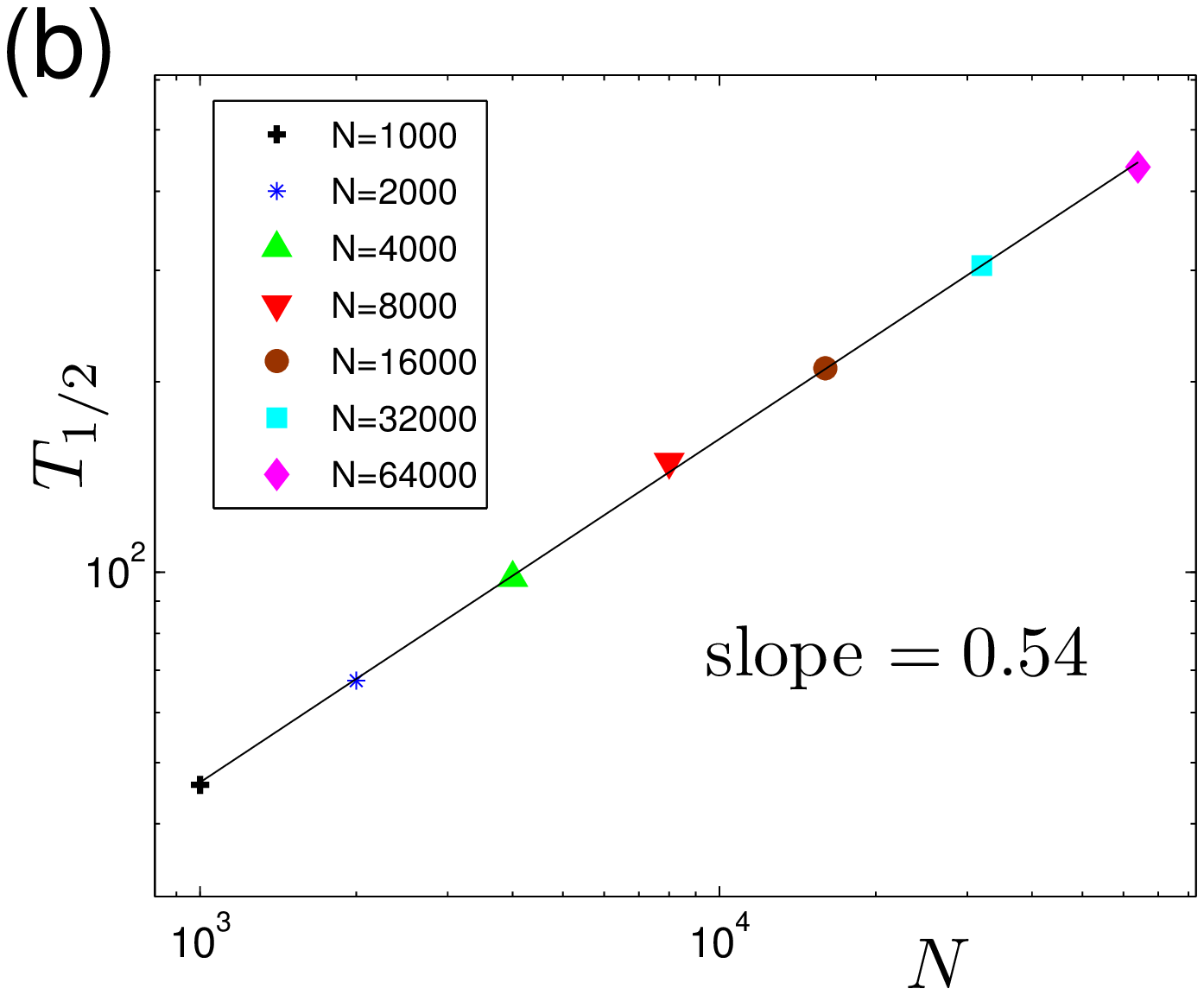}
 \includegraphics[width=4.1cm]{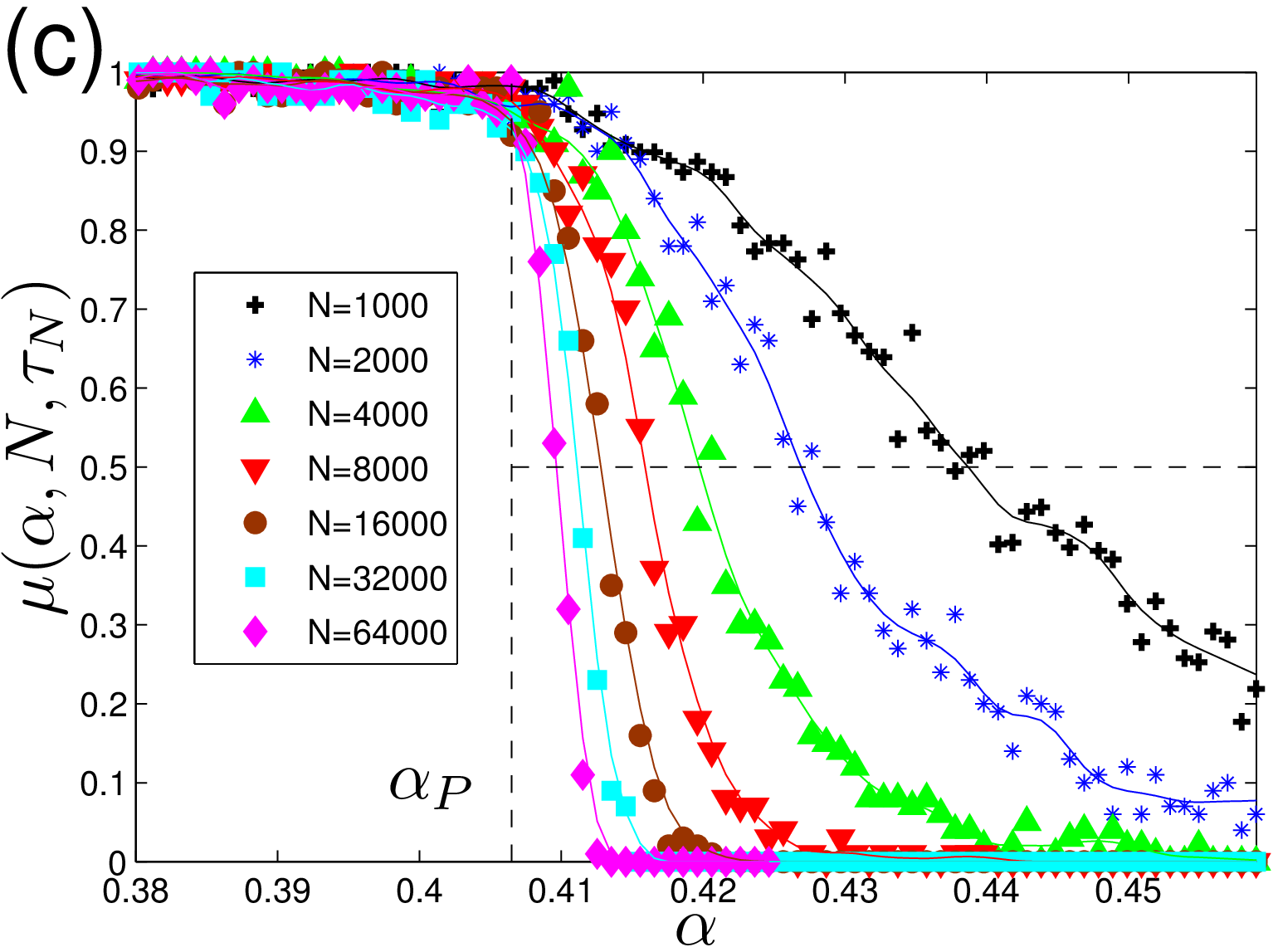}
 \includegraphics[width=4.1cm]{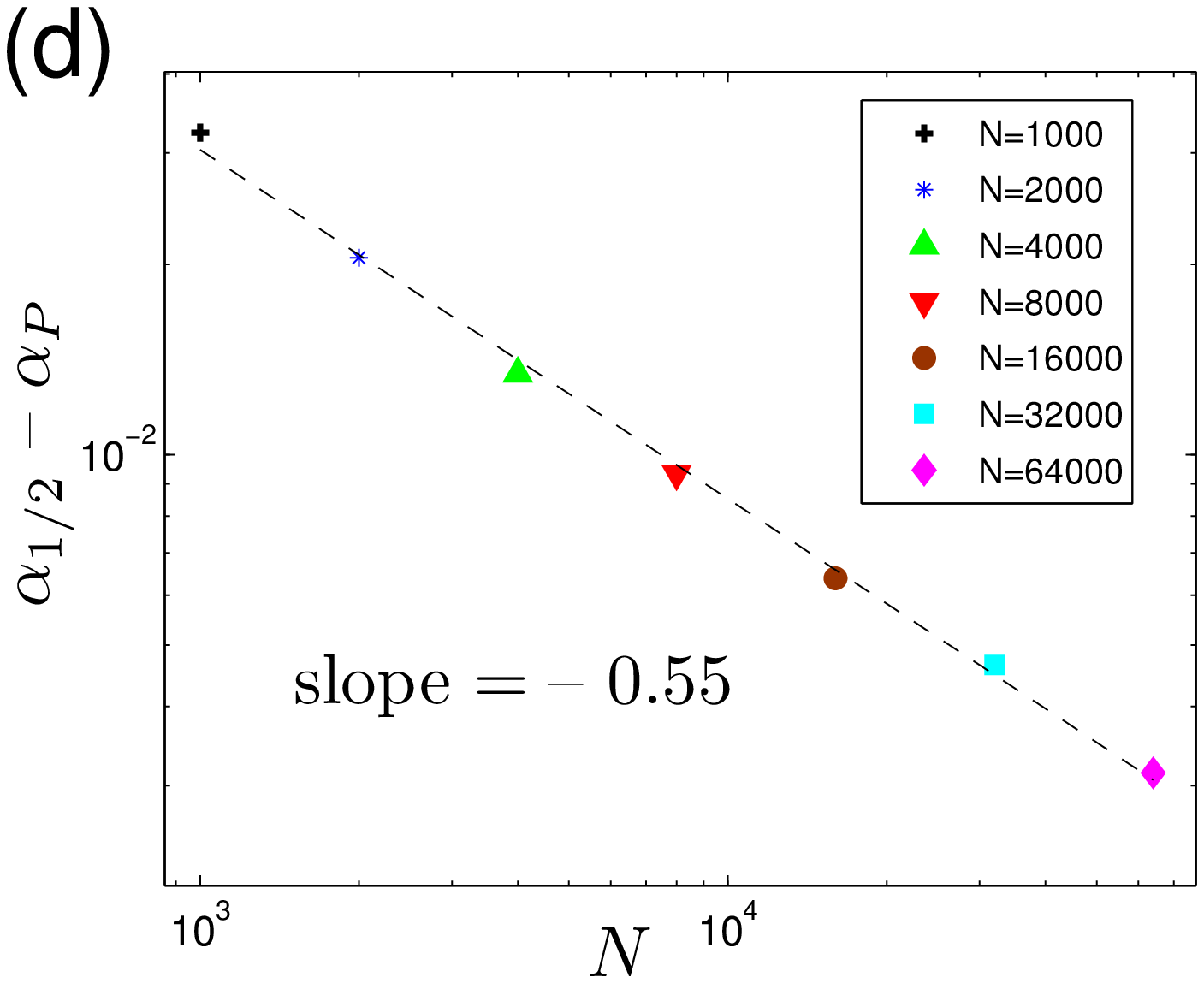}
\caption{\small Finite size scaling analysis of the nonequilibrium transition from partial to complete synchronization at $\alpha_{\textnormal{P}}$ for $\sigma=1.0$. In (a) simulation runs with system sizes up to $N=10^6$ were performed. At the critical point complete synchronization merges with the metastable partially synchronized state which is approached as $t^{-1}$ if the transition is of mean field directed percolation universality. The line $(1-r(t))\sim t^{-1}$ is drawn for comparison. The inset shows the linear approach of the mean order parameter $r\to 1$ in the vicinity of the critical point. The line of slope $1.01$ is a linear fit to the data in double-logarithmic scales. Other critical exponents are obtained from the time statistics for a realization to reach the absorbing state of complete synchronization. Figure (b) shows the median of this time at the critical point for various system sizes. In (c) we plot the fraction $\mu_0$ of 100 realizations which reach complete synchronization before the time $\tau_N = 1900 (N/64000)^{0.54}$ for different $N$ as a function of $\alpha$. This defines a median $\alpha_{1/2}$, shown in (d), which approaches the critical point $\alpha_{\textnormal{P}}=0.4065$ at a power law with exponent $-0.55$ as a function of $N$.}
\label{Fig:FiniteScaling}
\end{figure}
In finite systems, the partially synchronized state may disappear after a
long transient time, and the state of complete synchronization is
approached at an exponential rate. This sudden change of behavior
resembles the transition from an active state to the absorbing state in
directed percolation processes \cite{DPreview00}.
\\ \\
Finite size scaling analysis in low-dimensional coupling topologies has demonstrated that
synchronization in coupled map lattices is of Kardar-Parisi-Zhang or
directed percolation universality \cite{AhlPiko02}. In a small world
network, the transition is expected to be of mean field universality \cite{OstilliMendes08,HongChoi02}.
\\ \\
To examine this point we have performed a finite size scaling analysis in the vicinity of $\alpha_{\textnormal{P}}$ at fixed shortcut density $\sigma=1.0$~.
The initial condition was a stable partially synchronized state at $\alpha=0.52$ and $R\approx 0.75$. After a transient time we decreased $\alpha$ and observed how $r(t)$ approached zero.
From simulation runs with single realizations of networks with sizes up to $N=10^6$ oscillators in the vicinity of the transition point, we found $\alpha_{\textnormal{P}} = 0.4065$ and $(1-R) \sim (\alpha-\alpha_{\textnormal{P}})^\beta$ with $\beta=1.01$ (Fig.~\ref{Fig:FiniteScaling}). This exponent is consistent with mean field universality of directed percolation.   
The nonequilibrium nature of the phase transition is expressed in the time dependence of the probability distribution of the order parameter. In a finite directed percolation system with a unique absorbing state, this absorbing state will be reached with probability one after a long enough transient time, even when a stable nonzero mean field solution exists \cite{GaveauSchu93}. 
\\ \\
A suitable absorbing
state for a network of coupled phase oscillators can be defined
via a Lyapunov function.
Let $V:[0,2\pi]^N \to [0,2\pi]$ denote the minimal length of the arc that contains the phases of all oscillators. Then one can show that $V$ is a Lyapunov function in any subset $B_S$ of the phase space with $V(B_S) \le \pi-2\alpha$.
If the network is strongly connected, i.e., there exists a directed path
between any two nodes, and $H_{ij}\ge 0$ then one can show that $\dot V<0$ for all $V>0$
in $B_S$. Thus $B_S$ defines an absorbing state in which complete synchronization, i.e., $V=0$, is approached exponentially. 
\\ \\
Let $\mu(\alpha,N,t)$ denote the probability of a process to have reached the absorbing state prior to the time $t$, then the finite size scaling ansatz for this probability is
\beq	\label{Eq:FSAnsatz}
	\mu(\alpha,N,t) = \tilde\mu\left((\alpha-\alpha_{\textnormal{P}})N^{\nu},N^{-z} t\right)	~.
\eeq
This probability can be determined by averaging over a large number of simulation runs. At each value of $\alpha$ in Fig.~\ref{Fig:FiniteScaling}c we have performed 100 simulations for system sizes $1000\le N\le 64000$.
To determine the exponent $z$ we have performed $1000$ simulation runs for each system size at the critical parameter $\alpha=\alpha_{\textnormal{P}}=0.4065$ and $\sigma=1.0$.
At the transition point the median $T_{1/2}$ defined as $\mu(\alpha,N,T_{1/2})=0.5$ scales with the system size as
\beq
	T_{1/2}(N) \sim N^z	~.
\eeq
The third critical exponent $\nu$ is related to the width and the position of the transition. Defining the position of the transition as $\mu(\alpha_{1/2},N,N^zT_0)=0.5$ one finds
\beq
	(\alpha_{1/2}-\alpha_{\textnormal{P}}) \sim N^{-\nu}	~.
\eeq
Confidence intervals for the experimentally determined values of $z$ and $\nu$ could be estimated by bootstrapping on the sample of simulation runs. However, we presume that systematic errors are dominating for the relatively small system sizes used in our simulations leading to slightly higher values of $z$ and $\nu$ than the expected value of $0.5$ for mean field directed percolation. We find $z\approx 0.54$ and $\nu\approx 0.55$.
%
%
\section*{V. Discussion}
So far, we have presented our numerical results. In this section,
we 
discuss the different dynamical regimes of our model and the mechanism for desynchronization.

\subsection*{Topological crossover}
Our small-world network model exhibits a topological cross-over between
low short cut densities $\sigma
< 1$ and high shortcut densities $\sigma > 1$.
The two regimes are characterized by the scaling of the average distance between successive joints of the network, which are oscillators that couple to more than one neighbor, along an arbitrary path (See Fig.~\ref{Fig:NetworkModel}).
By choosing the end points of the shortcuts randomly and uncorrelated, the number $L$ of nodes between two joints on the original
ring lattice is exponentially distributed as
$p(L)\sim\exp(-\sigma L)$ in the limit $N\to\infty$.
The expected
value of $L$ is $\langle L \rangle = \left(\exp(\sigma)-1\right)^{-1}$,
which is approximately $\sigma^{-1}$ at low shortcut densities and
$\exp(-\sigma)$ at high shortcut densities. This cross-over is reflected in
the phase dynamics of the incoherent state as a regime of traveling
waves on the original ring lattice at $\sigma<1$ to a regime without
clear spatio-temporal patterns at $\sigma>1$ (Fig.~\ref{Fig:phase_snapshot}).
\\ \\
The Poissonian degree distribution with a mean in-degree of $k=\sigma+1$
and standard deviation $\sqrt{k}$ provides homogeneity at high
shortcut densities. On the other hand, at low shortcut densities
$\sigma<1$, the network consists of one dimensional chain segments
which interact nonlinearly at the joints of the network. The network of
joints, where chain segments are replaced by edges and indirect
interactions, is also very homogeneous, since the number of joints which
receive exactly two inputs is of order $O(\sigma)$ while the total
number of all other joints is of order $O(\sigma^2)$. In this sense, the
network is most inhomogeneous at intermediate shortcut densities
$\sigma\approx 1$ where the amount of short chain segments and joints with more than two input links are comparable.
Indeed, this topological complexity is reflected in a maximum of the variance of the phase velocities at intermediate shortcut densities, 
which quantifies the strength of the internal noise (See Fig.~\ref{Fig:structural_resonance}). 
We will discuss the dynamical properties of our model for low and high
shortcut densities, separately.
\subsection*{Statistical properties of the incoherent state}
\begin{figure}[!t] 
\center
 \includegraphics[width=4.1cm]{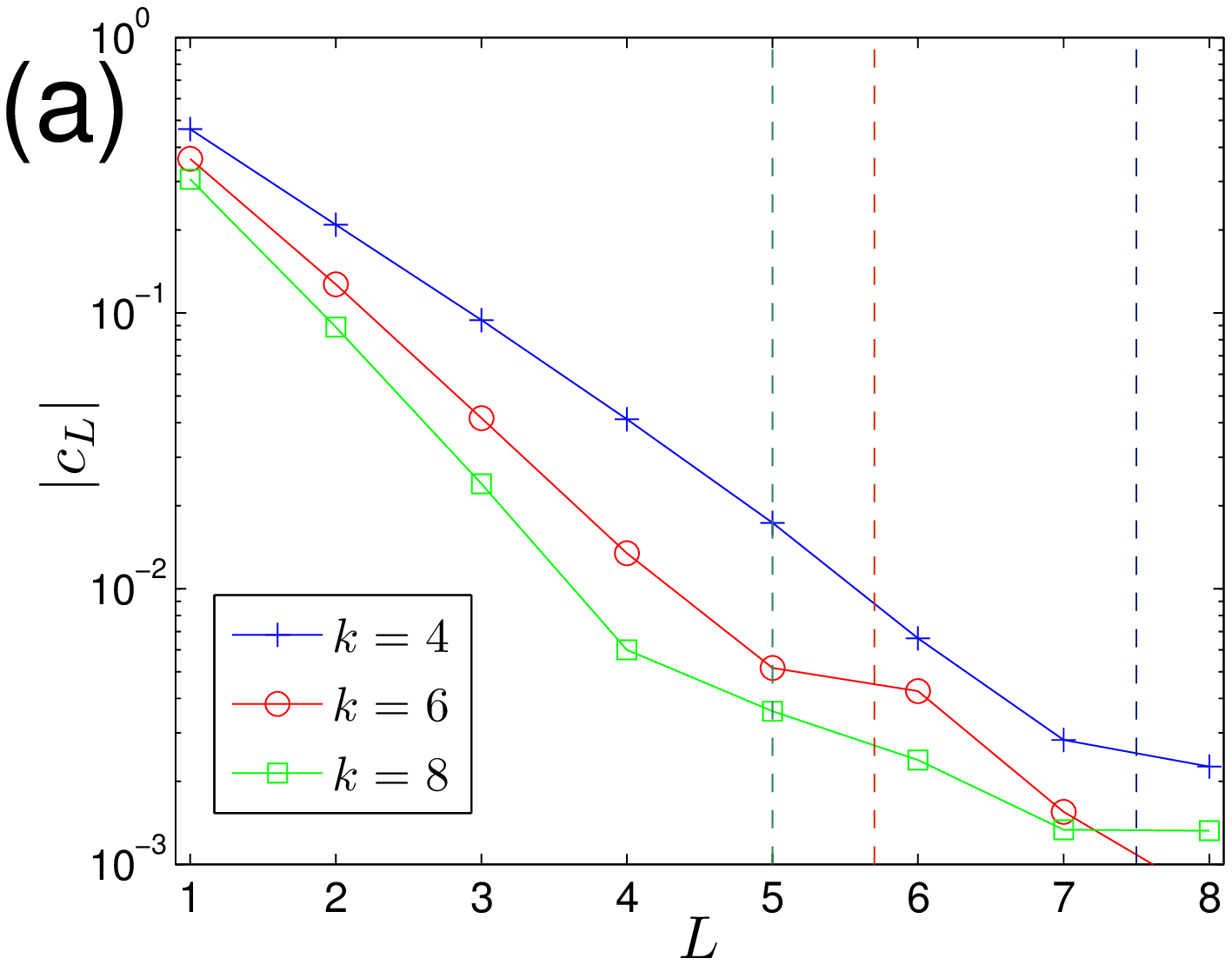}
 \includegraphics[width=4.1cm]{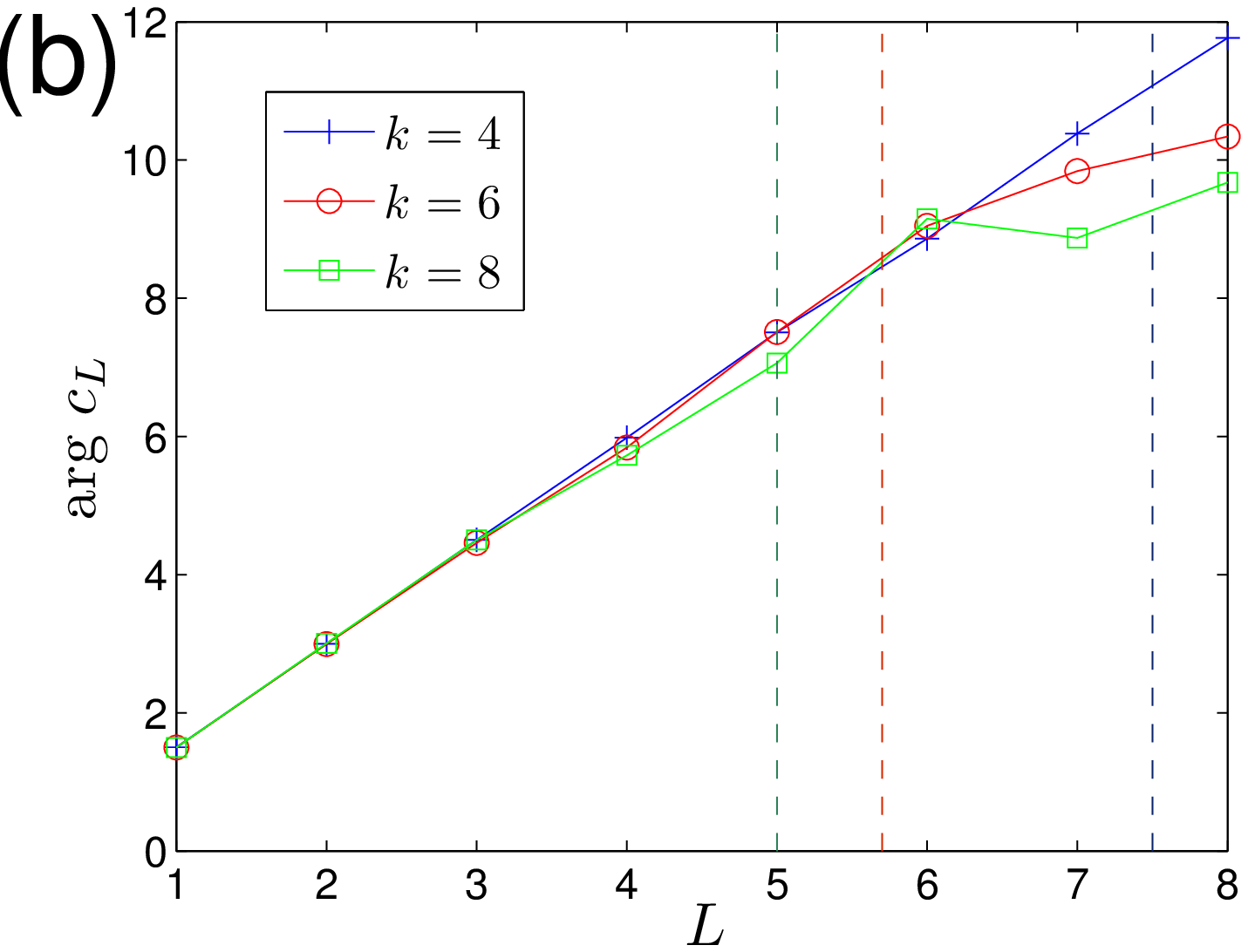}
 \includegraphics[width=4.1cm]{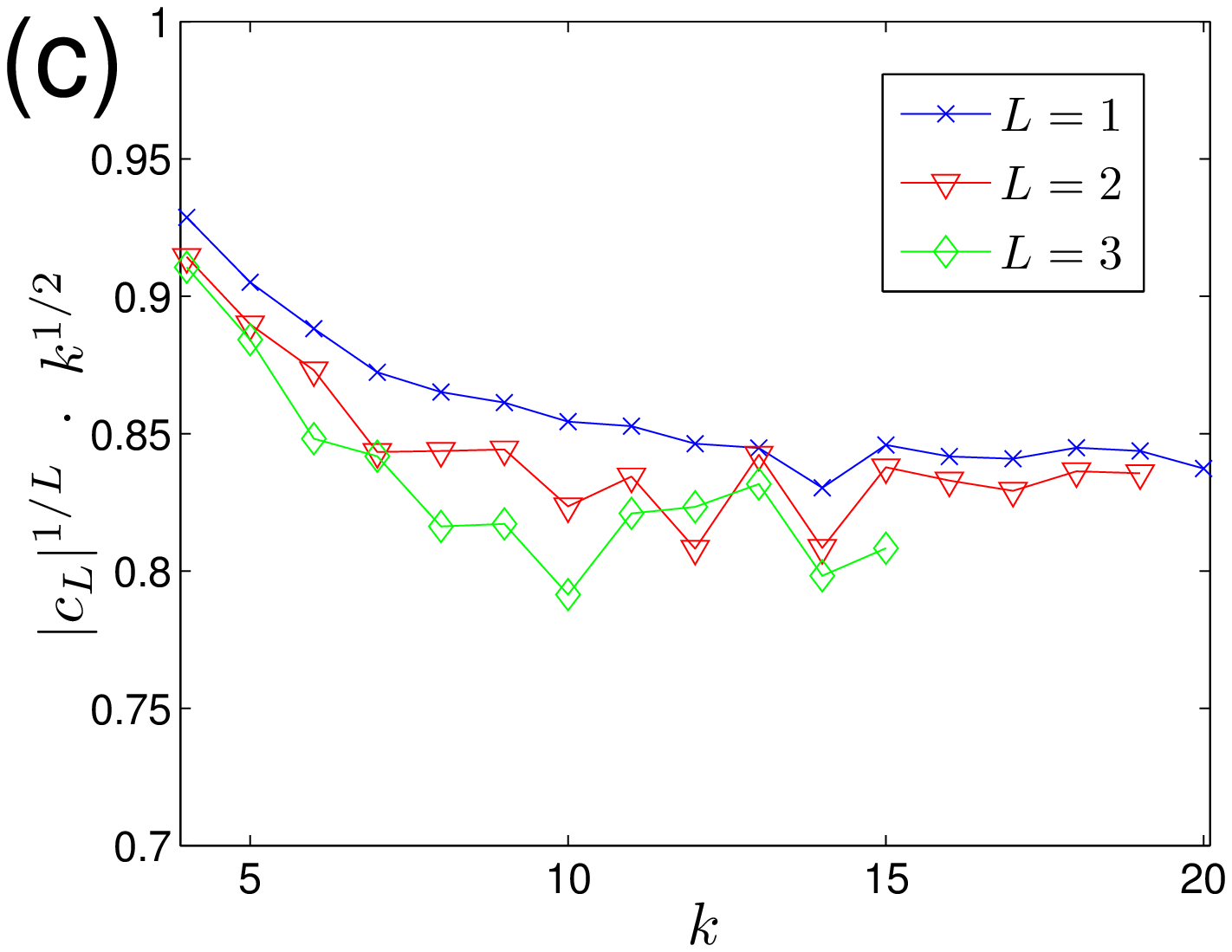}
 \includegraphics[width=4.1cm]{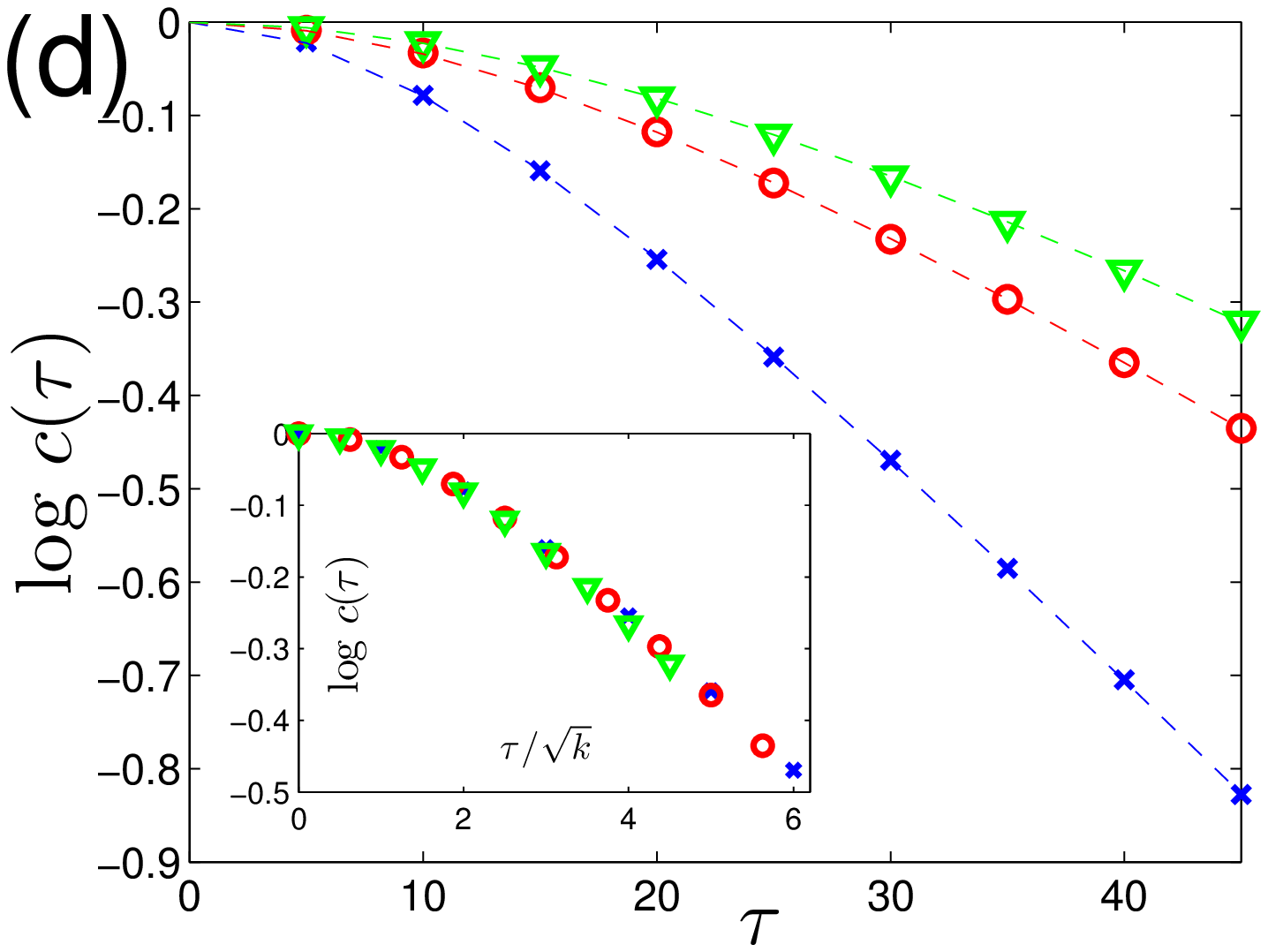}
\caption{\small Complex correlation function in the incoherent state. (a) Absolute value in logarithmic scales as a function of the distance $L$ on the ring backbone of the network, for $k=\sigma+1=4$ (blue crosses), $k=6$ (red circles) and $k=8$ (green squares). (b) Angle as a function of the distance $L$ on the ring backbone of the network for $k=4$ (blue crosses), $k=6$ (red circles) and $k=8$ (green squares). We used $\alpha=1.5$, well above the synchronization threshold. The dashed lines in subfigures (a) and (b) mark the mean distance $\log N / \log k$ in the network of size $N=32000$. (c) Scaling of absolute value with the number of neighbors at distance $L=1$ (blue crosses), $L=2$ (red triangles) and $L=3$ (green diamonds). (d) Logarithm of autocorrelation function at time difference $\tau$ for $k=25$ (blue crosses), $k=64$ (red circles) and $k=100$ (green triangles) and parametric fit to autocorrelation function of Brownian flight on the circle (Eq.~(\ref{Eq:BrownCorr}), dashed lines). See Table I for values of $D$ and $\gamma$. The inset shows the collapse of the curves under a rescaling of time with factor $k^{-1/2}$.}
\label{Fig:correlfun}
\end{figure}
In the incoherent state, the phases undergo a chaotic phase diffusion process. An oscillator coupled to a finite number of neighbors is therefore subject to a fluctuating force. In the incoherent state in a unidirectional network, the neighbors of an oscillator have independent phases, because their distance is of the order of the network diameter. Thus, if the number of neighbors $k$ is sufficiently large, the local mean fields will fluctuate around zero with approximately Gaussian distribution and one can show that its variance is equal to $k^{-1}$. Since the amplitudes of the local mean fields determine the phase velocities of the oscillators which in turn generate the local mean fields, the chaotic phase diffusion process is invariant under a rescaling of time as $k^{-1/2}$ (See Appendix A for details). 
From the circular autocorrelation function $c(\tau)=\langle\cos(\vartheta(t+\tau)-\vartheta(t))\rangle$ we can estimate the effective phase diffusion constant $D$ and an effective scattering rate $\gamma$ of the chaotic phase diffusion process by comparing it directly to the autocorrelation function of a Brownian flight on the circle \cite{Risken89} with
\beq	\label{Eq:BrownCorr}
	c(\tau) = e^{-D\tau + D\gamma^{-1}\left(1-e^{-\gamma\tau}\right)}	.
\eeq
The variance of the phase velocities of such a Brownian flight is equal to $v^2=D\gamma$ and can directly be compared to the variance $\textnormal{var}(\dot\vartheta)$ of the chaotic phase diffusion process.  In Table \ref{Tab:IncoStats} we show the experimentally determined effective diffusion constant and the variance of the phase velocities for $k=25,64$ and $100$, and $\alpha=\pi/2$.  We find the predicted scaling of $D$, $\gamma$ and $\textnormal{var}(\dot\vartheta)$. In particular the rescaled effective phase diffusion constant is $D\sqrt{k}\approx 0.13$.
\\ \\
Note that these statistical properties do not
depend on $\alpha$  
as can be seen in Fig.~\ref{Fig:sigma_alpha_scans}. 
%
\begin{table}[!t]
\center
 	\begin{tabular}{|c|r|r|r|}
		\hline
			$k$						&	25	&	64	&	100	\\
		\hline
			$D\sqrt{k}$					&	0.1252	&	0.1296	&	0.1417	\\
			$\gamma\sqrt{k}$				&	0.4064	&	0.3937	&	0.3598	\\
			$k \textnormal{var}(\dot\vartheta)$ 		&	0.0478	&	0.0519	&	0.0544	\\
			$k D \gamma = k v^2$ 				&	0.0509	&	0.0510	&	0.0510	\\
			$\alpha_{\textnormal{HP}}$			&	1.27	&	1.39	&	1.43	\\
		\hline
	\end{tabular}
\caption{\small Timescales of the chaotic phase diffusion process for various large mean degrees $k\gg 1$. We find that the effective phase diffusion constant and the effective scattering rate scale with $1/\sqrt{k}$ and the variance of the phase velocities scales as $1/k$ (See Fig.~\ref{Fig:structural_resonance}b). The transition point to synchronization $\alpha_{\textnormal{HP}}$ has been determined with the help of our control scheme.}
\label{Tab:IncoStats}
\end{table}
\subsection*{Synchronization transition for high shortcut density}
In the incoherent state each oscillator in the network is subject to a fluctuating local mean field generated by a finite number of neighbors performing the chaotic phase diffusion process. These neighbors sample from the global distribution of phases. While the global mean field evolves deterministically in the thermodynamic limit $N\to\infty$, the local mean fields will be approximately distributed as a Gaussian around the global mean field with a variance of order $O(1/k)$, a relaxation rate $\gamma$ and a diffusion constant $D$ of order $O(1/\sqrt{k})$. 
%
%
Let us define the complex correlation function
\beq
	c_{ij}	=	\langle z_i^* z_j \rangle
\eeq
where $z_j=\exp(\textnormal{i}\vartheta_j)$ and the average is over time. In our homogeneous network model, the correlation is a function $c_L$ of the length $L$ of the shortest directed path from $j$ to $i$. A heuristic mean field ansatz is to use this correlation function as a distance dependent weight and phase shift for the coupling
\beq	\label{Eq:MF01}
	{\dot\vartheta}_i ~=~ R ~ \frac{\sum_{L=1}^\infty \textnormal{Im}\left[ c_L^* e^{\textnormal{i}(\Theta-\vartheta_i)}\right] }{\sum_{L=1}^\infty |c_L|}
\eeq
where $R$ and  $\Theta$ are the mean field amplitude and angle. Here we have chosen a co-rotating reference system where $\langle\dot\vartheta\rangle=0$. In the incoherent state the complex correlation function (Fig.~\ref{Fig:correlfun}) has the form 
\beq
	c_L ~=~ \varrho^{|L|}  e^{\textnormal{i} L\alpha} ~=~ \left(\varrho_0  k^{-\frac{1}{2}}\right)^{|L|} e^{\textnormal{i}L\alpha}
\eeq
Each shell, i.e., a set of oscillators $j$ with constant distance $L$ from $i$, is correlated to the oscillator $i$ with an amplitude that decreases exponentially with $L$ and at an angle which changes linearly with $L$. The factor $\varrho_0$ was determined to be approximately $0.85$~(See Fig.~\ref{Fig:correlfun}c). We are now able to express the phase evolution as that of a globally coupled system with effective coupling strength $\kappa_{\textnormal{eff}}$ and effective asymmetry $\alpha_{\textnormal{eff}}>\alpha$ as
\beq	\label{Eq:MF02}
	{\dot \vartheta}_i ~=~ \kappa_{\textnormal{eff}} ~R~ \sin\left(\Theta - \vartheta_i - \alpha_{\textnormal{eff}} \right) ,
\eeq
where $\kappa_{\textnormal{eff}}$ and $\alpha_{\textnormal{eff}}$ are determined by
\beq
	\kappa_{\textnormal{eff}} ~e^{-\textnormal{i}\alpha_{\textnormal{eff}}} ~=~ \frac{(1-\varrho)}{\varrho} \sum_{L=1}^\infty \left(\varrho e^{-\textnormal{i}\alpha}\right)^L ~=~ \frac{(1-\varrho)e^{-\textnormal{i}\alpha}}{1-\varrho e^{-\textnormal{i}\alpha}}	.
\eeq
This mean field ansatz predicts a stable incoherent state whenever $\alpha_{\textnormal{eff}}\ge \pi/2$ which is fullfilled if
\beq
	\cos\alpha < \varrho_0 k^{-\frac{1}{2}}	.
\eeq
The critical line $\alpha_{\textnormal{HP}}=\arccos(\varrho_0 / \sqrt{k})$ obtained from this heuristic mean field ansatz shows the same qualitative behavior as the critical line obtained from the control scheme, although the numerical values do not agree well (See Fig.~\ref{Fig:sa_bif}d). It predicts an asymptotic approach of $\alpha$ to $\pi/2$ at the order of $O(1/\sqrt{k})$ as $k\to\infty$.
\\ \\
Another heuristic argument can be made for the transition line $\alpha_{\textnormal{P}}(\sigma)$. 
Suppose that all oscillators have the same phase $\vartheta_i = 0$
except one oscillator with phase $\vartheta_0$ which is externally
driven. This oscillator assumes the role of an active seed in terms of percolation processes.
If this active seed activates (i.e., induces oscillations of) the other
oscillators that couple to the seed, the whole system may eventually
become active, resulting in the incoherent or partially synchronized states. We are thus
interested in the conditions under which an oscillator with phase $\vartheta_1$
and in-degree $k$ that couples to the seed can become active. 
The phase equation for $\vartheta_1$ is initially
\beq	\label{Eq:SeedPropDyn}
	{\dot\vartheta}_1 ~=~ \frac{1}{k}\sin\left(\vartheta_0 - \vartheta_1 - \alpha\right) ~-~ \frac{k-1}{k} \sin\left(\vartheta_1+\alpha\right) ~+~ \sin\alpha	~.
\eeq
At criticality, we expect that 
the phase difference $\vartheta_0-\vartheta_1$ is drifting fast so that the first term of
Eq.~(\ref{Eq:SeedPropDyn}) can be time-averaged and neglected. In this case, oscillator $1$ gets activated for $\sin\alpha > (k-1)/k$, and with $k = \sigma+1$ we obtain the condition
\beq	\label{Eq:AsinCond02}
	\alpha=\arcsin\frac{\sigma}{\sigma+1}
\eeq
for criticality. Equation (\ref{Eq:AsinCond02}) agrees surprisingly well
with the numerically determined transition line $\alpha_{\textnormal{HP}}(\sigma)$ for $\sigma>1$ but less well with $\alpha_{\textnormal{P}}(\sigma)$ (Fig.~\ref{Fig:sa_bif}c,d). Again we predict that the critical $\alpha$ approaches $\pi/2$ at the order of $O(1/\sqrt{k})$.

\subsection*{Synchronization transition for low shortcut density}
For low shortcut densities $\sigma<1$, the situation becomes even more
complicated. Due to the topological cross-over another length scale $\sigma^{-1}$ is introduced in the system.
The dynamics in the incoherent state for low
shortcut densities is characterized by traveling waves along one
dimensional chain segments of the original ring topology which interact
nonlinearly at the joints of the network (See
Fig.~\ref{Fig:phase_snapshot}).  We call the first and last oscillators
in a chain segment the head and the tail of the chain, respectively.
For normalized input strength the chain segments are always phase locked
to the head but act as a low pass filter to the phase dynamics (Appendix C). As a
result, the variance of the phase velocities decreases along a chain segment and for smaller values of $\sigma$ (Fig.~\ref{Fig:structural_resonance}).  
The interaction between the joints of the network is indirect and involves delay and additional phase shifts.
Also, the correlation between the head and an oscillator down the chain segment decreases exponentially with the
distance between the head and the oscillator. These factors seem to inhibit synchronization so that the critical $\alpha_{HP}(\sigma)$ is still decreasing for $\sigma\ll 1$ but much slower than linearly (Fig.~\ref{Fig:sa_bif}a) .

\section*{VI. Summary and Conclusions}
We have shown that a finite average number of neighbors and an asymmetry of the phase coupling function can inhibit synchronization in
homogeneous networks of identical oscillators. Using a
control technique, we could construct the bifurcation diagram for the order
parameter.
\\ \\
A Finite size scaling analysis at the nonequilibrium phase transition from partial synchronization to complete synchronization shows critical exponents of mean field universality.
\\ \\
In contrast to \cite{Ermentrout08} it is not the heterogeneity of the
node degree distribution that drives the 
system away from synchronization. Instead, it is the interplay between the spatial structure of the network and the internal noise which prevents the oscillators from reaching synchronization.
The temporal fluctuations generated by the system itself in the chaotic incoherent or partially synchronized state give rise to a correlation function that decays exponentially with the distance. Using this correlation function to formulate a heuristic mean-field theory for the incoherent state, we have qualitatively explained the transition from incoherence to synchronization.
\\ \\
Our main results are derived from numerical simulations. An analytic
description of the transition curve and the chaotic phase diffusion process in the incoherent state remain challenging open problems.
\section*{Acknowledgment}
We thank Hugues Chat{\'e} and Kazumasa Takeuchi for valuable discussion about directed percolation processes.
R.T. acknowledges Prof. Yasumasa Nishiura, funding through a JSPS short term fellowship PE.07606 and by JST Special Coordination Funds for Promoting Science and Technology. N.M. acknowledges the support through the Grants-in-Aid for Scientific Research (No. 20760258 and 20540382) from MEXT, Japan.
%
%

%
%
\section*{Appendix A : Scaling Argument for $\sigma\gg 1$}
Here we will present the details of the scaling argument that relates
the amplitude and the time-scale of the fluctuations in the local
fields.  Let us consider 
the dynamics of the complex state $z_i=\exp(\textnormal{i}\vartheta_i)$ of an oscillator
\beq	\label{Eq:ComplexKPE}
	{\dot z}_i	~=~	\textnormal{i}~\textnormal{Im}\left[z_i^*\Psi_i e^{-\textnormal{i}\alpha}\right]~z_i		~=~	r_i ~\textnormal{i}~\textnormal{Im}\left[e^{\textnormal{i}(\Theta_i-\theta_i-\alpha)}\right] ~z_i	~,
\eeq
where $\Psi_i$ is the local field given by 
\beq
	\Psi_i	=	\sum_j H_{ij} z_j	= r_i e^{\textnormal{i}\Theta_i}.
\eeq
The amplitude and the time scale of
local field dynamics can be inferred from the complex correlation
function
\beq	\label{Eq:PsiCorr}
	c_\Psi(\tau)	=	\left\langle \Psi_i^*(t+\tau)\Psi_i(t) \right\rangle	~,
\eeq
where $\left\langle\dots\right\rangle$ indicates the
time average with respect to the stationary processes $z_j(t)$. Assuming
$k$ independent neighbors with correlation function
\beq
	c_z(\tau)	=	\left\langle z_j^*(t+\tau)z_j(t) \right\rangle
\eeq
one finds exactly
\beq	\label{Eq:Equal_Corr}
	c_\Psi(\tau) 	=	\frac{1}{k} c_z(\tau)		~.
\eeq
This relation is remarkable since it is valid for all $k$ and all
time differences $\tau$. Therefore, the phase dynamics of the local fields has
exactly the same time scale as the dynamics of the phase oscillators.
Without calling on the central limit theorem this relation also gives
the mean square amplitude of the local fields as $\left\langle r^2
\right\rangle = c_{\Psi}(0) = k^{-1}$ for all $k$.  
\\ \\
The assumption of independent phases of the neighbors only holds in large unidirectional networks in the incoherent state. Due to the chaotic dynamics, correlations between oscillators decay exponentially with the distance in the network which is of order $O(\log N / \log k)$ (Fig.~\ref{Fig:correlfun}).
\\ \\
The effective phase diffusion constant can be defined by the asymptotic behavior of
$c_\Psi(\tau)$ and $c_z(\tau)$ at large $\tau$
\beq	\label{Eq:DeffDef}
	D = -\lim_{\tau\to\infty}\frac{d}{d\tau} \ln c(\tau)	~.
\eeq
We recall that the circular autocorrelation function for a free Brownian flight on the circle is
$c(\tau) = \exp\left(-D\tau + D/\gamma \left( 1 - \exp(-\gamma\tau)\right)\right)$, where $\gamma$ is an effective scattering rate and $D$ is the effective phase diffusion constant \cite{Risken89}. Exponential decay is expected at time scales $\tau\gamma\gg 1$.
From Eqs.~(\ref{Eq:Equal_Corr}) and (\ref{Eq:DeffDef}) it follows that the
effective phase diffusion constants of the single oscillator phase and
that of the local fields are identical. For sufficiently large $k$, the
local fields are normally distributed and it is only the amplitude of
the local fields that decreases when $k$ is increased. The process
(\ref{Eq:ComplexKPE}) is therefore invariant under a rescaling
\beqarr
	t'		&=&		\sqrt{\frac{k}{k'}}t			\nonumber \\ \\
	\Psi'(t')	&=& 		\sqrt{\frac{k}{k'}}\Psi(t)	~.	\nonumber
\eeqarr
In our model, this rescaling is achieved by changing the number of
neighbors from $k$ to $k'$. The second transformation affects both the
amplitude and the phase diffusion timescale of the local fields. 
\\ \\
Because of the homogeneity of our model, we can assume statistical
equivalence of the phase dynamics for all oscillators.  The phase
$\vartheta_i(t)$ of an oscillator that is coupled via
Eq.~(\ref{Eq:ComplexKPE}) to $k$ independent but identically distributed
phase diffusion processes $\vartheta_j(t)$ must be another realization
of the same process. In particular, the effective phase diffusion
constant must be the same.  To get insight into the phase
diffusion process, let us examine an oscillator with complex state $z$
that is coupled to a complex Ornstein-Uhlenbeck process $\Psi$
\beqarr \label{Eq:OUdrift} {\dot\Psi} &=& - D \Psi + \sqrt{D}\eta
	\nonumber \\ \\ {\dot z} &=& \textnormal{i}z
	\textnormal{Im}\left[z^*\Psi\right] \nonumber \eeqarr
where $\eta$ is complex Gaussian white noise of unit strength in real and imaginary part.
\begin{figure}[tb] 
\center
 \includegraphics[width=9.0cm]{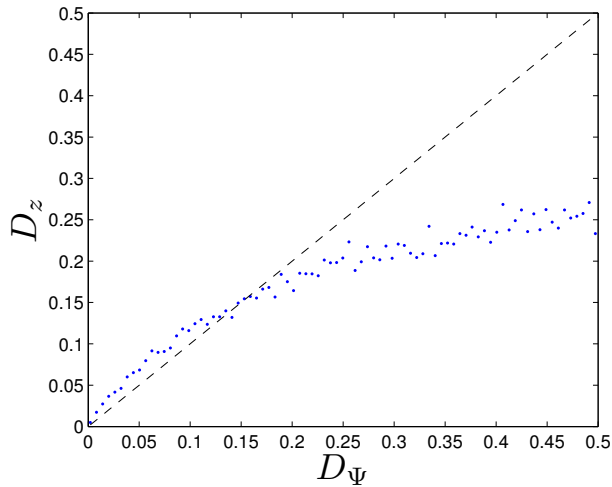}
\caption{\small Effective phase diffusion constant
$D_z$ of a phase oscillator with complex state
variable $z$ coupled to a complex valued Ornstein-Uhlenbeck process of
unit variance and phase diffusion constant $D_\Psi$
(Eq.~(\ref{Eq:OUdrift})). The fixed point at $D^*\approx 0.15$ is
expected to be close to the rescaled effective phase diffusion constant
$D(k)\sqrt{k}$ of the characteristic stationary phase
diffusion process in the incoherent state for $k\gg 1$. 
}  \label{Fig:DinDout}
\end{figure}
Note that the Ornstein-Uhlenbeck process $\Psi$ is only an approximation of the dynamics of an actual local field which has the same Gaussian stationary distribution
with expected square amplitude $\left\langle r^2 \right\rangle = 1$. The effective
phase diffusion constant as defined in Eqs.~(\ref{Eq:PsiCorr}) and (\ref{Eq:DeffDef}) is
$D_\Psi=D$ \cite{Risken89}. The effective phase
diffusion constant $D_z$ depends nonlinearly on
$D_{\Psi}$. For large values of $D$ the time scales of
$\Psi$ and $z$ are separated and the phase of $z$ will diffuse much
slower than the phase of $\Psi$. For very small values of $D$ the phase
of $z$ will almost always be locked to the phase of $\Psi$. Only when
$\Psi$ diffuses near zero, phase slips may occur which add a ballistic
component to the evolution of $z$ and can increase the effective phase
diffusion constant $D_z$ to a value larger than
$D$. 
\\ \\
There exists a fixed point $D^*$ for which the time scale of $\Psi$ and $z$ are identical. 
The fixed point $D^*$ should be close to the actual value of the
self-consistent solution. A self-consistent solution must also yield
Eq.~(\ref{Eq:Equal_Corr}). The average field of $k$ such processes $z$
may be normally distributed but is not an Ornstein-Uhlenbeck
process. However, the iterative procedure of coupling a phase oscillator
to $k$ independent neighbors defines a mapping in the space of
stationary processes on the circle. Starting with a Brownian phase
diffusion we see that in the first iteration the map is contracting with
respect to the effective diffusion constant (Fig.~\ref{Fig:DinDout}), an
indication for the existence of a unique stationary phase diffusion
process characterizing the incoherent state.  
\\ \\ 
We have measured
$D_z$ as a function of $D$ in numerical integration of
Eq.~(\ref{Eq:OUdrift}) and found the the fixed point to be $D^* \approx
0.15$. Table \ref{Tab:IncoStats} shows the effective diffusion constants in our model for $k=25, 64$ and $100$, obtained from a nonlinear parametric least square fit to the autocorrelation function (system average). We find that $D\sqrt{k}\approx 0.13$ agrees well with this value.
\section*{Appendix B : Control Scheme}
Given the mean field dynamics
\beq
	\dot r = f(r,\alpha)
\eeq
of a scalar order parameter $r$ with a system parameter $\alpha$ we are interested in the bifurcation curve $\alpha_0(r)$ with $f(r,\alpha_0(r))=0$. If we can control the system parameter $\alpha = \alpha(t)$ then the linear control scheme
\beq
	\dot \alpha = c_0 (r-r_0) + c_1 \dot r
\eeq
has the fixed point $(r,\alpha)=(r_0,\alpha_0(r_0))$. With $f_0=\partial_r f(r_0,\alpha_0(r_0))$ and $f_1=\partial_\alpha f(r_0,\alpha_0(r_0))$ the Jaccobian of the combined mean field and control dynamics reads
\beq
	\mathit{J} = \left(
 	\begin{array}{cc}
 		f_0  & f_1 \\
 		c_0  & c_1 f_1
 	\end{array}
 	\right)		~.
\eeq
Assuming $f_1<0$, the conditions for stability $\textnormal{tr}(J)<0$ and $\textnormal{Det}(J)>0$ yield
\beq
	c_1 > -\frac{f_0}{f_1}	\quad\textnormal{and}\quad c_0 > f_0 c_1	.
\eeq
The assumption of $f_1<0$ applies, since an increase of $\alpha$ counteracts synchronization and decreases $r$ if $\alpha>\alpha_0(r)$.
Sufficient conditions to stabilize any point $(r_0,\alpha_0(r_0))$, regardless of the sign of $f_0$, i.e., of the stability of the uncontrolled fixed point, are
\beq
	c_0 > |f_0| c_1 > \frac{f_0^2}{|f_1|}	.
\eeq
\section*{Appendix C : The overdamped linear chain}
For low shortcut densities one can view the dynamics in the small-world
network as traveling waves which interact through a network of
joints. Each joint of the network is the head of a unidirectional chain
segment. A joint receives input from at least two nodes of other chain
segments. To understand the role of the chain segments in the
transition to synchronization, we study the dynamics of the phases
and the signal transmission along a chain segment in a linear
approximation.
\\ \\
The phase equations for a unidirectionally coupled chain of oscillators
are
\beq	\label{Eq:1dKPEs}
	{\dot\vartheta}_n	=	\sin\left(\vartheta_{n-1} -\vartheta_n - \alpha\right)	+ \sin\alpha~.
\eeq
We identify $\vartheta_0$ with the phase in the head oscillator of the
chain and $\vartheta_L$ with the phase of the tail oscillator.
Under appropriate boundary conditions Eq.~(\ref{Eq:1dKPEs}) allows for
traveling wave solutions $\dot\vartheta=\sin\alpha$ and $\vartheta_{n-1}
- \vartheta_n = \alpha$. In principle, all frequencies in the interval
$[\sin\alpha- 1,\sin\alpha+1]$ and corresponding phase differences are
possible but we choose the traveling wave solution with the average
frequency $\sin\alpha$. Small deviations $x_n$ from this solution can be
studied in a linear approximation
\beq
	{\dot x}_n	=	x_{n-1} - x_n	,
\eeq
which can be solved given the trajectory $x_0(t)$ in the head of the
chain. The system is asymptotically independent from initial conditions and the solution is
\beq
	x_n(t) = \int\limits_0^\infty e^{-\tau} x_{n-1}(t-\tau) d\tau	=	\int\limits_0^\infty \tau^{n-1} \frac{e^{-\tau}}{(n-1)!} ~ x_0(t-\tau) d\tau	~.
\eeq
The dynamics in the tail of a chain is a time convolution of the
dynamics in the head of the same chain around the traveling wave
solution. A discontinuous jump of the phase in the head of the chain
will translate at unit speed while the width of the phase jump grows at
the same rate, as a result of the gamma distribution for the time
convolution kernel. In this linear approximation, the deviation from the
traveling wave solution grows diffusively with the distance from the
head of the chain. Suppose the head of the chain performs a Brownian
motion with $\left\langle (x_0(t)-x_0(0))^2 \right\rangle = 2Dt$. Then
we obtain
\beq
	\left\langle(x_n(t)-x_0(t))^2\right\rangle = \left\langle x_n(t)^2\right\rangle_{x_0(t)=0},
\eeq
and
\beqarr
	\left\langle x_n(t)^2\right\rangle_{x_0(t)=0} &=& \left\langle x_n(0)^2\right\rangle_{x_0(0)=0}	\nonumber \\ \nonumber \\
	&=& \int\limits_0^\infty dT dT'	\frac{(TT')^{n-1}e^{-(T+T')}}{\Gamma^2(n)} \nonumber \\ &&\times\left\langle x_0(-T)x_0(-T')\right\rangle_{|x_0(0)=0}		\nonumber \\ \nonumber \\
	&=& 2\int\limits_0^\infty dT d\tau \frac{(T(T+\tau))^{n-1}e^{-T}e^{-(T+\tau)}}{\Gamma^2(n)} \nonumber \\ && \times \left\langle x_0^2(-T)\right\rangle_{|x_0(0)=0} \qquad	\nonumber \\ \nonumber \\
	&=&  4 D \int\limits_0^\infty dT \frac{T^{n}e^{-T}}{\Gamma(n)}\frac{\Gamma(n,T)}{\Gamma(n)}	~.
\eeqarr
Using
\beq
	\Gamma(n+1,T) = n\Gamma(n,T) + T^n e^{-T}	\quad,
\eeq
we obtain
\beqarr
	\left\langle x_n^2(0)\right\rangle_{|x_0(0)=0}	
	&=&  4 D \int\limits_0^\infty dT \frac{T^{n}e^{-T}}{\Gamma(n+1)}\frac{n}{\Gamma(n+1)} \nonumber \\ &&\times \left(\Gamma(n+1,T)-T^ne^{-T}\right) \qquad	\nonumber \\ \nonumber \\
	&=&  2 D n \left(1 - 2\int\limits_0^\infty \frac{T^{2n}e^{-2T}}{\Gamma^2(n+1)}\right)	\nonumber \\ \nonumber \\
	&=&  2 D n \left(1 - \frac{\Gamma(2n+1)}{\Gamma^2(n+1)}4^{-n}\right) \nonumber \\ \nonumber \\ &\approx& 2 D n \left(1 - \frac{1}{\sqrt{n\pi}}\right)	~.
\eeqarr
The effective spatial phase diffusion constant along a linear chain of
oscillators is the same as the temporal diffusion constant of the phase
in the head of the chain. The complex correlation function
\beq
	c_{L0}(0) = \left\langle z_L^*(t)z_0(t) \right\rangle \sim e^{i\alpha L - DL}	~\quad\textnormal{for}\quad L\gg 1
\eeq
has an amplitude that decreases exponentially with the chain length. A lower shortcut density therefore decreases the correlation between the joints of the network, making it more difficult to synchronize.
\newpage

\begin{thebibliography}{10}

\bibitem{OsChanKu07}
G.~V. Osipov, J.~Kurths, and C.~Zhou,
\newblock {\em Synchronization in oscillatory networks},
\newblock Springer Series in Synergetics, Springer, Berlin, 2007.

\bibitem{PiRoKurths01}
A.~Pikovsky, M.~Rosenblum, and J.~Kurths,
\newblock {\em Synchronization : A universal concept in nonlinear sciences},
  volume~12 of {\em Cambridge Nonlinear Science Series},
\newblock Cambridge University Press, Cambridge, 2001.

\bibitem{Winfree80}
A.~T. Winfree,
\newblock {\em The geometry of biological time}, volume~12 of {\em
  Interdisciplinary Applied Mathematics},
\newblock Springer-Verlag, New York, second edition, 2001.

\bibitem{Kuramoto84}
Y.~Kuramoto,
\newblock {\em Chemical oscillations, waves, and turbulence}, volume~19 of {\em
  Springer Series in Synergetics},
\newblock Springer-Verlag, Berlin, 1984.

\bibitem{DiAre08}
A.~D\'{\i}az-Guilera and A.~Arenas,
\newblock {\em Bio-Inspired Computing and Communication}, pp.184., Springer-Verlag, Berlin, 2008.

\bibitem{LaKoPeHe06}
S.~L\"ammer, H.~Kori, K.~Peters, and D.~Helbing,
\newblock Physica A: Statistical Mechanics and its Applications {\bf 363}, 39
  (2006),
\newblock Information and Material Flows in Complex Networks.

\bibitem{SiFaWie93}
M.~Silber, L.~Fabiny, and K.~Wiesenfeld,
\newblock J. Opt. Soc. Am. B {\bf 10}, 1121 (1993).

\bibitem{Kuramoto75}
Y.~Kuramoto,
\newblock Self-entrainment of a population of coupled non-linear oscillators,
\newblock in {\em International {S}ymposium on {M}athematical {P}roblems in
  {T}heoretical {P}hysics ({K}yoto {U}niv., {K}yoto, 1975)}, pages 420--422.
  Lecture Notes in Phys., 39, Springer, Berlin, 1975.

\bibitem{RespOtt05}
J.~G. Restrepo, E.~Ott, and B.~R. Hunt,
\newblock Phys. Rev. E (3) {\bf 71}, 036151, 12 (2005).

\bibitem{OttAnton08}
E.~Ott and T.~M. Antonsen,
\newblock Chaos {\bf 18}, 037113, 6 (2008).

\bibitem{PikRos08}
A.~Pikovsky and M.~Rosenblum,
\newblock Phys. D {\bf 238}, 27 (2009).

\bibitem{Ermentrout08}
T.-W. Ko and G.~B. Ermentrout,
\newblock Phys. Rev. E (3) {\bf 78}, 026210, 8 (2008).

\bibitem{WattsStrogatz98}
D.~J. Watts and S.~H. Strogatz,
\newblock Nature {\bf 393}, 440 (1998).

\bibitem{NewmanWatts99}
M.~E.~J. Newman and D.~J. Watts,
\newblock Phys. Rev. E {\bf 60}, 7332 (1999).

\bibitem{OstilliMendes08}
M.~Ostilli and J.~F.~F. Mendes,
\newblock Phys. Rev. E (3) {\bf 78}, 031102, 26 (2008).

\bibitem{BarratWeigt00}
A.~Barrat and M.~Weigt,
\newblock The European Physical Journal B - Condensed Matter and Complex
  Systems {\bf 13}, 547 (2000).

\bibitem{DoGoMen02}
S.~N. Dorogovtsev, A.~V. Goltsev, and J.~F.~F. Mendes,
\newblock Phys. Rev. E {\bf 66}, 016104 (2002).

\bibitem{HongChoi02}
H.~Hong, M.~Y. Choi, and B.~J. Kim,
\newblock Phys. Rev. E {\bf 65}, 026139 (2002).

\bibitem{Derrida86}
B.~Derrida and Y.~Pomeau,
\newblock EPL (Europhysics Letters) {\bf 1}, 45 (1986).

\bibitem{KlemmBorn05}
K.~Klemm and S.~Bornholdt,
\newblock Phys. Rev. E (3) {\bf 72}, 055101, 4 (2005).

\bibitem{Roxin04}
A.~Roxin, H.~Riecke, and S.~A. Solla,
\newblock Phys. Rev. Lett. {\bf 92}, 198101 (2004).

\bibitem{Timme08}
S.~Jahnke, R.-M. Memmesheimer, and M.~Timme,
\newblock Phys. Rev. Lett. {\bf 100}, 048102 (2008).

\bibitem{KissHudson05}
I.~Z. Kiss, Y.~Zhai, and J.~L. Hudson,
\newblock Phys. Rev. Lett. {\bf 94}, 248301 (2005).

\bibitem{BlaToe05}
B.~Blasius and R.~T\"onjes,
\newblock Phys. Rev. Lett. {\bf 95}, 084101 (2005).

\bibitem{KuraBot02}
Y.~Kuramoto and D.~Battogtokh,
\newblock Nonlinear Phenomena in Complex Systems {\bf 5}, 380 (2002).

\bibitem{AbraStrog04}
D.~M. Abrams and S.~H. Strogatz,
\newblock Phys. Rev. Lett. {\bf 93}, 174102 (2004).

\bibitem{BlaToe09}
R.~T\"onjes and B.~Blasius,
\newblock Phys. Rev. E {\bf 80}, 026202 (2009).

\bibitem{BaraPecora02}
M.~Barahona and L.~M. Pecora,
\newblock Phys. Rev. Lett. {\bf 89}, 054101 (2002).

\bibitem{DPreview00}
H.~Hinrichsen,
\newblock Advances in Physics {\bf 49}, 815 (2000).

\bibitem{AhlPiko02}
V.~Ahlers and A.~Pikovsky,
\newblock Phys. Rev. Lett. {\bf 88}, 254101 (2002).

\bibitem{GaveauSchu93}
B.~Gaveau and L.~S. Schulman,
\newblock J. Statist. Phys. {\bf 70}, 613 (1993).

\bibitem{Risken89}
H.~Risken,
\newblock {\em The {F}okker-{P}lanck equation}, volume~18 of {\em Springer
  Series in Synergetics},
\newblock Springer-Verlag, Berlin, second edition, 1989,
\newblock Methods of solution and applications.


\end{thebibliography}
\end{document}